\documentclass[fleqn,usenatbib]{mnras}

\usepackage{newtxtext,newtxmath}

\usepackage[T1]{fontenc}

\DeclareRobustCommand{\VAN}[3]{#2}
\let\VANthebibliography\thebibliography
\def\thebibliography{\DeclareRobustCommand{\VAN}[3]{##3}\VANthebibliography}

\usepackage{graphicx}	
\usepackage{amsmath}	
\usepackage{pdflscape}

\title[Sparse regression modeling of stellar mass]{A sparse regression approach to modeling the relation between galaxy stellar masses and their host halos}

\author[M. Icaza-Lizaola et al.]{
M. Icaza-Lizaola$^{1,2}$\thanks{E-mail: miguel.a.de-icaza-lizaola@durham.ac.uk},
Richard G. Bower$^{1,2}$, 
Peder Norberg$^{1,2,3}$, 
Shaun Cole$^{1}$,
\newauthor
Matthieu Schaller$^{4,5}$,
Stefan Egan$^{6}$
\\
\scriptsize $^{1}$Institute for Computational Cosmology, Department of Physics, University of Durham, South Road, Durham DH1 3LE, UK.\vspace*{-2pt} \\
\scriptsize $^{2}$Institute for Data Science, Department of Physics, University of Durham, South Road, Durham DH1 3LE, UK.\vspace*{-2pt} \\
\scriptsize $^{3}$Centre for Extragalactic Astronomy, Department of Physics, University of Durham, South Road, Durham DH1 3LE, UK.\vspace*{-2pt} \\
\scriptsize $^{4}$Lorentz Institute for Theoretical Physics, Leiden University, PO Box 9506, NL-2300 RA Leiden, The Netherlands\\
\scriptsize$^{5}$Leiden Observatory, Leiden University, PO Box 9513, NL-2300 RA Leiden, The Netherlands\\
\scriptsize $^{6}$Procter $\&$ Gamble, Newcastle Innovation Centre, Newcastle upon Tyne, UK.\vspace*{-2pt} \\
}

\date{Accepted XXX. Received YYY; in original form ZZZ}

\pubyear{2020}

\begin{document}
\label{firstpage}
\pagerange{\pageref{firstpage}--\pageref{lastpage}}
\maketitle

\begin{abstract}
{Sparse regression algorithms have been proposed as the appropriate framework to model the governing equations of a system from data, without needing prior knowledge of the underlying physics. In this work, we use sparse regression to build an accurate and explainable model of the stellar mass of central galaxies given properties of their host dark matter (DM) halo. Our data set comprises 9,521 central galaxies from the EAGLE hydrodynamic simulation. By matching the host halos to a DM-only simulation, we collect the halo mass and specific angular momentum at present time and for their main progenitors in 10 redshift bins from $z=0$ to $z=4$. The principal component of our governing equation is a third-order polynomial of the host halo mass, which models the stellar-mass halo-mass relation. The scatter about this relation is driven by the halo mass evolution and is captured by second and third-order correlations of the halo mass evolution with the present halo mass. An advantage of sparse regression approaches is that unnecessary terms are removed. Although we include information on halo specific angular momentum, these parameters are discarded by our methodology. This suggests that halo angular momentum has little connection to galaxy formation efficiency. Our model has a root mean square error (RMSE) of $0.167 \log_{10}(M^*/M_\odot)$, and accurately reproduces both the stellar mass function and central galaxy correlation function of EAGLE. The methodology appears to be an encouraging approach for populating the halos of DM-only simulations with galaxies, and we discuss the next steps that are required.}
\end{abstract}




\section{Introduction}

Gravitational collapse in the expanding Universe leads to the formation of complex, highly non-linear structures.  The force of gravity can be accurately modeled through N-body cosmological simulations. However, observational probes of the Universe's structure usually rely on galaxies, bringing into play a much broader range of baryon physics.  Unlike dark matter only (DMO) simulations, which only allow interactions through gravity, baryon simulations need to deal with complicated feedback processes and are strongly influenced by events happening at scales much smaller than the size of the simulation grid \citep{10.1111/j.1365-2966.2009.16029.x}. While this can be mitigated by including sub-grid models of these processes, in the form of sources or sinks of energy and matter, the resulting computational cost of accurate baryonic simulations remains far greater than that of DMO simulations. As a consequence the volume of the universe that can be modelled in this way is limited. A hybrid approach is therefore necessary, in which a large volume DMO simulation is populated with galaxies based on the halo-galaxy relationship found in smaller baryonic simulations.  This requires a methodology that can extract robust halo-galaxy relationships making optimal use of the available volume of baryonic simulations. In this paper, we explore whether sparse-regression models, which are a type of machine learning algorithm, provide an attractive approach.

A full reconstruction of the baryonic Universe would require us to also model satellite galaxies. These are subject to additional physics such as tidal striping, heating \citep{1983ApJ...264...24M} and other environmental processes. In this work, we focus on developing and presenting our methodology, applying it to model central galaxies. We leave the extension of our methodology to include satellite galaxies for a future work.

It is already well established that there is a strong correlation between the stellar mass ($M^*$)  of a central galaxy and the mass of its host halo \citep{10.1093/mnras/183.3.341}. This relation is known as the Stellar Mass -- Halo Mass (SMHM) relation. However, there is a significant scatter in the SMHM relation \citep[e.g.][]{10.1111/j.1365-2966.2010.17436.x,10.1093/mnras/stv2062} which indicates that the stellar mass of a galaxy may also depend on other factors. Here, we investigate whether the formation history and the angular momentum of the host dark matter halo also play a role. Dependence on formation history is often referred to as assembly bias \citep{2004MNRAS.350.1385S,2005MNRAS.363L..66G,Gao_2007,10.1093/mnras/stz2344}. The effect of assembly bias in the EAGLE simulation has been studied in \cite{10.1093/mnras/stw1225}, where it was concluded that it might alter the clustering signal amplitude of the sample by up to $20\%$. It is worth noting, however, that while assembly bias has been detected in several simulations, the efforts made to detect it on observations have been inconclusive to date \citep[e.g.][]{Lin_2016,Tojeiro2017,2020arXiv201004176S}.

To explore the effect of assembly bias, \cite{Matthee_2016} studied the correlation between the residuals in the SMHM relation and different DM properties on EAGLE. They found that the parameters that are most correlated with this residual are those that are determined by the evolution of the halo mass, in particular concentration and halo formation time. They found no other parameter strongly correlated with the residual of the SMHM relation once it was corrected for the halo concentration correlation. Our aim in this paper is to investigate the optimal measure of halo formation trajectory and to determine whether the prediction of stellar mass can be improved by including the additional halo specific angular momentum. 

In observations, the angular momentum of a galaxy appears correlated with its stellar mass \citep{Fall_2013}. However, while there is a correlation between the history of the specific angular momentum of a galaxy and its host halo \citep{10.1093/mnras/stw1286}, \cite{Danovich_2015} uses cosmological simulations to suggest that the specific angular momentum of gas and dark matter undergo decoupled formation histories and that it is only the final distribution of spin parameters that is similar between baryons and dark matter. Nevertheless, it remains physically  plausible that halo specific angular momentum and galaxy formation efficiency maybe interconnected in some more complex way.

The aim of this work is to develop a sparse regression approach to find a polynomial equation that relates the stellar mass of a galaxy with the properties of its DM halo. This is a form of Machine Learning (ML). More conventional ML algorithms such as neural networks \citep[e.g.][]{2015Natur.521..436L} and random forests \citep[e.g.][]{2001MachL..45....5B} are some of the most powerful tools for parameterising a data set. However, algorithms like neural networks or ensemble models \citep{roberts_everson_2001} generate models with virtually no explainability, so that extracting the physics behind the model would be difficult. While the network could predict galaxy properties, it would be hard to gain confidence that the output is physically reasonable. Random forest algorithms work by building a collection of decision trees that are designed to be as uncorrelated as possible. These models are easier to interpret, as one can measure how often a variable was used and how drastically the entropy decreases in each step. A potential issue with some machine learning algorithms is the slow evaluation speed of a final model. 

Sparse regression methods \citep[SRM;][]{10.5555/2834535} are a set of minimization algorithms that are efficient at discarding unnecessary free parameters. This makes them very useful at minimizing functions for which one suspects that most free parameters are irrelevant except for a small subset that one is trying to identify. SRM provide a trade-off between including very many free parameters (which would result in over-fitting to random artefacts in the data) and eliminating too many parameters (which would result in a poor description of the data). SRM have been proposed as the appropriate framework to extract the governing equations of a physical system from the data alone with relatively little prior knowledge required of the system's physics \citep{Brunton3932}. A key advantage of the SRM approach is that the small number of retained coefficients are likely to have a clearer physical interpretation. Further more, given that the models produced with SRM can be simple polynomial equations, their evaluation comes with virtually no computational cost.
 
We apply the sparse regression methodology to model the stellar mass of galaxies in the EAGLE 100~Mpc hydrodynamical simulation \citep{2015MNRAS.446..521S,2015MNRAS.450.1937C}. The EAGLE simulation provides a reasonable description of the observed Universe \citep{2015MNRAS.450.1937C,10.1093/mnras/stx1263,2015MNRAS.450.4486F,10.1093/mnras/stw1230}, and is ideal for our proposes as parameter values of the DM halos are stored at several redshift slices \citep{2016A&C....15...72M}, which permit us to model assembly bias. While all models presented here were calibrated on EAGLE data, we expect that the methodology can be applied to other hydrodynamical simulations with similar success. One common issue with this type of analysis is the danger of including a selection bias in the independent variables due to dark matter halos in hydrodynamical simulations being affected by baryonic processes that might alter properties like their density profile \citep[e.g.][]{10.1093/mnras/stv1341,10.1111/j.1365-2966.2012.20879.x,1996MNRAS.283L..72N}. With this in mind, we use a one-to-one matching \citep{10.1093/mnras/stv1067} between our hydrodynamical simulation and a dark matter-only simulation built using the same properties and initial conditions.

Our work builds on other ML methods that have shown promising developments in the creation of mock catalogs using DM halos. \cite{Moster_2021} uses neural networks to populate DM halos from N-body simulations with galaxies. While their goal is similar to ours, the philosophy behind both models is different. Their approach avoids using hydrodynamical simulations and focuses on placing the galaxies inside halos in such a way that it reproduces observed properties of the galaxy populations. While that approach leads to accurate models, it would by construction be hard to extract any physical interpretation out of it. \cite{10.1093/mnras/sty1719} used a random forest algorithm to predict which DM particle in a simulation would end up inside a DM halo of a given mass, while \cite{2019MNRAS.482.2861B} used a neural network to build DM halo mocks. 

In this paper, we focus on the properties of central galaxies. For mock simulations to be compared to observations of large-scale structure surveys they need to be populated with galaxies in such a way that they reproduce the stellar mass function (SMF) and the clustering patterns of galaxies. This would require us to assign both a central and a population of satellite galaxies to each dark matter halo. We discuss the additional challenges of modelling the stellar mass of satellite galaxies at the end of the paper. 
 
This paper is organized as follows. Section~\ref{section:Sparse_Regresion} introduces the sparse regression methodology used to build our model and includes an example model that illustrates the behavior of the algorithm. Section~\ref{DataSet} introduces the hydrodynamical simulation from which the input data were extracted and discusses how the data was processed to be used by our algorithm. The details on running the algorithm using the data set are presented in  Section~\ref{Runing_alghoritm}. As sections~\ref{section:Sparse_Regresion} and \ref{Runing_alghoritm} introduce and test our methodology, readers primarily interested in the astrophysical results can go directly to section \ref{results}. Section~\ref{results} shows the results of the different configurations in which we run our code, and we discuss the physical interpretation of different terms of our governing equations and compare the stellar mass distribution and clustering statistics to those from EAGLE. Our conclusions and final thoughts, along with a brief discussion on the next steps that we aim to take, are presented in Section~\ref{conclusions}.

\section{The Sparse Regression Methodology}
\label{section:Sparse_Regresion}

This section starts by setting the general problem in \S\ref{problem_statment}. This is followed, in \S\ref{sparse_regression}, by an introduction to the sparse regression method considered, i.e.\ the Least Absolute Shrinkage and Selection Operator (LASSO). We explain our minimization implementation in \S\ref{Minimisation} and the penalty hyperparameter definition in \S\ref{Penalty}. We end in \S\ref{subsection:toy_model} with a simple example to more clearly illustrate our methodology.
 
\subsection{Problem statement}
\label{problem_statment}
We are interested in finding a function that models a physical property, $y'$, that might be determined by a set of M variables $\vec{x}'=[x'_1,....,x'_M]$ (we reserve the symbol $x$ for normalised variables - see below). In this work $y'$ is the stellar mass of a galaxy and $\vec{x}'$ a set of present and past properties of its DM halo. We can build a data set of values of $y'$ and their associated $\vec{x}'$ by looking at large catalogues where the value of both has been measured. In this paper we use the output of the EAGLE hydrodynamical simulations (see Section~\ref{DataSet}).

We collect a sample of N galaxies to build a vector $\vec{y}'$, where
\begin{equation}
\vec{y}' = [y'_1,...,y'_N],
\end{equation}
\noindent and an associated matrix $\mathbf{X'}$, with each row $\vec{x}'_\alpha$ ($1 \le \alpha \le N$) representing the list of dependent variables associated with the DM halo of the corresponding  galaxy $y'_\alpha$:
\begin{equation}
\label{X_matrix}
\mathbf{X'}=
\begin{bmatrix}
   \vec{x}'_1 \\
   .\\
   \vec{x}'_N
\end{bmatrix}
=
\begin{bmatrix}
   x'_{11}&... & x'_{1M} \\
   .&.&.\\
   x'_{N1}&... & x'_{NM}
\end{bmatrix}
\end{equation}
The different columns of matrix $\mathbf{X'}$ correspond to different properties of the DM halo, where each property 
can have different units and distributions. It is, therefore, necessary to standardize our data. We choose to do this using the mean and standard deviation of the distribution, and define the normalised variable as:
\begin{equation}
\label{Normalisation}
{\vec{z}_i}=\frac{{\vec{z}'_i}-\mu({\vec{z}'_i})}{\sigma({\vec{z}'_i})},
\end{equation}

where $\vec{z}'_i$ is now a column of $\mathbf{X'}$ and $1 \le i \le M$ and  $\mu$ and $\sigma$ are the mean and standard deviation operators. The same normalization scheme is also applied to our dependent variable: $y=(y'-\mu(y'))/\sigma(y')$. Note that the primed variables refer to natural quantities and non-primed variables to standardised ones that have zero mean and unit variance.

The observed values of the M variables of $\vec{x}_\alpha$ will be used as inputs for a series of functions whose output one hopes to use to predict $y_\alpha$. These functions can in principle have any desired form, and so we will use a gradient descent algorithm to fit a linear combination of them to $\vec{y}$ (Section~\ref{Minimisation}). Although other approaches like singular value decomposition \cite{SingValueDesc} could be used in the hyperbolic case, we wish to ensure that the method is generic.

We consider a library of $D$ functions, and their evaluated values for the observed parameters of the $\alpha^{th}$ galaxy  $\vec{f}(\vec{x}_\alpha)=[f_1(\vec{x}_\alpha),.....,f_D(\vec{x}_\alpha)]$. The library of functions that we use in this work consists of:

\begin{itemize}

    \item A constant term $f^0(\vec{x}_\alpha)=1$.
    
    \item $M$ linear terms of the form [$f^1_1(\vec{x}_{\alpha})$, ..., $f^1_M(\vec{x}_{\alpha})$] = [$x_{\alpha 1}$, ..., $x_{\alpha i}$, ..., $x_{\alpha M}$] where $1 \leq i \leq M$.
    
    \item $M(M+1)/2$ quadratic terms of the form  [$f^2_1(\vec{x}_{\alpha})$, ..., $f^2_{M(M+1)/2}(\vec{x}_{\alpha})$] = [$x_{\alpha 1}^2$, ...., $x_{\alpha i}\,x_{\alpha j}$, ..., $x_{\alpha M}^2$] with $1 \leq i \leq j \leq M$. 
    
    \item $M(M+1)(M+2)/6$ cubic terms of the form
    [$f^3_1(\vec{x}_{\alpha})$, ..., $f^3_{M(M+1)(M+2)/6}(\vec{x}_{\alpha})$] = [$x_{\alpha 1}^3$, ... , $x_{\alpha i}\,x_{\alpha j}\,x_{\alpha k}$, ...., $x_{\alpha M}^3$] with $1 \leq i \leq j \leq k \leq M$. 

\end{itemize}

The total number of functions considered is:
 \begin{equation}
 \label{N_function}
 D= 1 +M + \frac{M(M+1)}{2} + \frac{M(M+1)(M+2)}{6} .
 \end{equation}
The number M of DM halo properties that we use depends on the specific parametrisation of the present and past properties of the halo that we select. We consider four different models each with different values of M (Section~\ref{Experiment_list}).

This methodology is able to deal with far more complicated functional forms than the polynomial forms used here. For example, we experimented with exponential decays  and step functions. However including these more complicated functions in our initial testing did not improve our models, but increased the computational time so  we excluded them from our final fits in this paper.

Our goal is to find optimized values of the coefficients $\vec{C}=[C_1,.......,C_D]$ that will make the linear combinations of our $D$ functions a sufficiently accurate model of $\vec{y}$. Specifically, we aim to find the optimized values of $\vec{C}$ such that $\vec{F}(\vec{C},\mathbf{X})\approx \vec{y}$, where $\vec{F}(\vec{C},\mathbf{X})$ is defined as:

\begin{equation}
\label{matix_definition}
\vec{F}(\vec{C},\mathbf{X})=\mathbf{F}(\mathbf{X})\vec{C}^T=
\begin{bmatrix}
  f_1({\vec{x}_1})&... & f_D({\vec{x}_1)} \\
   \vdots & \vdots & \vdots\\
  f_1({\vec{x}_N})&... & f_D({\vec{x}_N)}
\end{bmatrix}
\begin{bmatrix}
  C_1 \\
   \vdots \\
   C_D
\end{bmatrix}.
\end{equation}
We discuss the precise meaning of the approximate equality in the following section. Our aim is to achieve a balance between the accuracy of fitting the data while keeping the model as simple as possible. Clearly there is an underlying assumption that the functions included can be linearly combined into a sufficiently accurate model. In the absence of a detailed understanding of the physical system our approach is to include a large number of functions in our library, spanning the possible range of physical interactions. 

\subsection{Sparse regression}
\label{sparse_regression}
 Sparse regression methods aim to minimise the error term
 $\vert \vec{F}(\vec{C},\mathbf{X})-\vec{y}\vert$ while discarding any unnecessary functions by setting their associated coefficients $C_j$ to a negligibly small value. This makes them the appropriate framework for our problem as it allows us to include a large number of functions while knowing that all of the unnecessary ones will be discarded by the methodology. The fewer surviving coefficients the easier it is to interpret the solution (i.e. the more  explainable it is). The solution will also be less susceptible to over-fitting to random fluctuations in the training data.

One of the most common sparse regression algorithms is LASSO \citep[Least Absolute Shrinkage and Selection Operator;][]{doi:10.1111/j.2517-6161.1996.tb02080.x,2017arXiv171200484T}, where one minimizes
\begin{equation}
\label{equationtomin}
L =\chi^2+\lambda P(\vec{C}).
\end{equation}
$P(\vec{C})$ is known as the penalty function and its value should increase with the absolute value and number of coefficients that are not set to zero. The coefficient $\lambda$ is a hyperparameter of the model and determines the relative magnitude of the penalty term.
The value of $\lambda$ is determined using a k-fold methodology \citep{10.5555/2834535}, as described in Section~\ref{lambda_set}.

$\chi^2$ is the normal chi-squared function defined as
\begin{equation}
\label{chi2}
\chi^2=\sum_{\alpha=1}^N\frac{(F_\alpha(\vec{C},\mathbf{X})-y_\alpha)^2}{\sigma_{y_\alpha}^2},
\end{equation}
where $\sigma_{y_{\alpha}}$ is an estimate of the uncertainty in measurement $y_\alpha$ and $F_\alpha(\vec{C},\mathbf{X})$ is the $\alpha^{th}$ element of $\vec{F}(\vec{C},\mathbf{X})$. In the absence of the penalty term, $L$ would be the negative of the logarithmic likelihood function (ie., $L = -2\ln\mathcal{L}$).

In the standard LASSO approach $P(\vec{C})$ is defined as
\begin{equation}
\label{LASSO}
P(\vec{C})=\sum_{i=1}^D \mid C_i \mid .
\end{equation}
We introduce a regularisation term to smooth out the gradient discontinuities that occur when parameters are close to zero,
\begin{equation}
\label{Intermediate}
P(\vec{C})=\sum_{i=1}^D \mid C_i \mid e^{-\left( {\epsilon}/{C_i} \right)^2},
\end{equation}
where $\epsilon$ is a small constant. Note that $\exp(-\left( {\epsilon}/{C_i} \right)^2 )$ is very close to zero when $|C_i|\ll \epsilon$ and close to one when $|C_i|\gg \epsilon$. Therefore $\epsilon$ determines how close to zero a coefficient $C_i$ needs to go before its contribution to the penalty is negligible. We adopt a value of $\epsilon=10^{-3}$, which we show in Section~\ref{Runing_alghoritm} makes unnecessary coefficients go close enough to zero to be clearly distinguished from the ones that are useful, while keeping a reasonable computational cost (the closer to zero unnecessary coefficients are required to get the longer the minimizer needs to run).  We define a cutoff value $\nu$ as the threshold between used and discarded parameters: every coefficient larger than $\nu$ will be used in our model and all smaller coefficients are discarded. The exact value of $\nu$ is presented in Section~\ref{Runing_alghoritm}.

In equation~\ref{Intermediate} the contribution of each coefficient $C_i$ is independent of the contribution of all other coefficients. This means that there is not a strong penalty for having many small, but larger than $\epsilon$, values of $C_i$. We found that a more efficient approach at eliminating non-essential coefficients is to consider the contribution of a coefficient, in comparison to all of the other surviving coefficients. This is achieved by the following penalty function $P(\vec{C})=\sum_{i=1}^D \left[\sum_{j\neq i} \mid C_j \mid \right]\mid C_i \mid$. Combining both modifications our penalty function has the following form
\begin{equation}
\label{Penalty}
P(\vec{C})=\sum_{i=1}^D  \left[\sum_{j\neq i} \mid C_j \mid e^{-\left( {\epsilon}/{C_j} \right)^2} \right] \mid C_i \mid e^{-\left( {\epsilon}/{C_i} \right)^2}.
\end{equation}
This is the form of the penalty function adopted in our algorithm.

The $\chi^2$ is a measure of the  goodness of fit, which decreases as the model becomes more accurate. Balancing of the goodness of fit statistic and penalty term makes sparse models robust against over-fitting: an over-fitted model would use many parameters to make a unrealistically good fit, which would make the $\chi^2$ very small but it would also make the penalty term large (as it grows with the number of parameters). The minimum should belong to a model that is as simple as possible, while still being a sufficiently good fit. This is why when using a large library of functions all but a small subset of the coefficients end up being set to zero.

By making some general assumptions, we can estimate that in the optimised solution $P(\vec C)= \mathcal{O}(1)$. First we note that $P(\vec{C}) \approx \sum_{i=1}^D  \left[\sum_{j\neq i} \mid C_i \mid \mid C_j \mid \right]\approx(\sum_{i=1}^D \mid C_i \mid)^2$, and that the optimised solution should satisfy $\vec{F}(\vec{C},\mathbf{X})=\mathbf{F}(\mathbf{X})\vec{C}^T \approx \vec{y}$ 

Secondly let us note that, in our case, $F_i(\mathbf{X})$ correspond to third order combinations of elements of $\vec x_i$, with each element standardised to be of the order of magnitude of the elements of $\vec y$ and therefore $F_{i\alpha}(\mathbf{X})\approx\mathcal{O} (y_\alpha)$. 
From here it should be that $\sum_{i=1}^D \mid C_i \mid \approx \mathcal{O}(1)$, and consequently that $P(\vec{C}) \approx \mathcal{O}(1)$.

The properties of the simulated galaxies do not have formal measurement errors, but we still expect a random scatter due to the stochastic nature of the formation process. We therefore estimate a constant $\sigma^2_{y} = \sigma^2_{y_\alpha}$ (for $1 \le \alpha \le N$) using
\begin{equation}
\label{sigmay}
    \sigma^2_{y}=\sum_{\alpha=1}^N\frac{(F_\alpha(\vec{C},\mathbf{X})-y_\alpha)^2}{N}
\end{equation}
evaluated at $\vec C$ that minimizes equation~\ref{chi2} when $\sigma^2_{y_\alpha}=N$. A consequence of using this definition of $\sigma^2_{y_\alpha}$ is that if we then minimise Eq.~\ref{equationtomin} with no penalty ($\lambda=0$)  we find
\begin{equation}
     L(\lambda=0)=N.
    \label{eq:norm}
\end{equation}
The optimised value of $\lambda$ should be such that $\chi^2$ and $\lambda P(\vec{C})$ are of comparable size. Given that by $P(\vec{C}) \approx \mathcal{O}(1)$, and that we constructed $\sigma^2_{y_\alpha}$ such that $\chi^2 \approx \mathcal{O}(N)$ then $\lambda\approx \mathcal{O}(N)$. This allows us to estimate the sizes of penalty that we should explore. 
 
 \subsection{Minimization}
\label{Minimisation}
We use a gradient decent algorithm to minimise Eq.~\ref{equationtomin}. The process starts at an initial point in parameter space and iteratively walks in the direction opposite to the gradient of $L$ with respect to $C_i$. We use a variation of \cite{garfken67:math}, the standard practice for most machine learning methodologies. The size of each step is determined by a parameter $\eta$. At every step one computes $L$, if it is larger at the new position then $\eta$ is reduced (as it would likely mean that it overshot the minimum). In the opposite case $\eta$ size is increased if $L$ is smaller at the new position as it is likely that we are still far from the minimum.

The gradient of  $L$ from Eq.~\ref{equationtomin} is computed with respect to the vector of coefficients $C_i$. In the standard methodology one makes a step in the direction of the gradient at the current position. However, we found that this did not perform well in the steep-sided valleys that characterise $L$. In such valleys, a step will overshoot the minimum, and as a consequence the next step would be in the opposite direction than the previous one but with a slightly smaller step size. Progress along the valley toward the global minimum is then slow. This makes convergence inefficient in high dimensional spaces, as the minimizer tends to jump from one wall of a potential well to the opposite wall at each step instead of following a more direct downwards path. 

We achieved performance gains by using the following adaptation of the algorithm for determining the next step of the minimization. 
Defining the position of the $i^{th
}$ step as  $p_i={C_1^i,...,C_D^i}$, the gradient vector
\begin{equation}
-\nabla (L) (p_i)=\frac{\partial L}{\partial C_1}(p_i),..,\frac{\partial L}{\partial C_D}(p_i) 
\end{equation}
points downhill towards the nearest local minimum.
Since we are only interested in the direction of the gradient and not its magnitude we can normalize the vector as 
\begin{equation}
\nabla \overline{L} (p_i)=
\nabla (L) (p_i) \Big/ \left\vert
\nabla (L) (p_i) \right\vert \, .
\end{equation}
We make a first trial step on the downhill direction that takes us to the following position in parameter space
\begin{equation}
   p_{i+1/2}=p_i-\eta \nabla (\overline{L}) (p_i). 
\end{equation}
The direction of the next step $p_{i+1}$ is given by the mean of the gradients at $p_i$ and $p_{i+1/2}$, 
\begin{equation}
p_{i+1}=p_i-\eta [  \nabla (\overline{L}) (p_i)+ \nabla(\overline{L}) (p_{i+1/2)}]/2  \, . 
\end{equation}
This swings the direction of travel to align with the valley.

In order to determine if our code has converged we look at the size of steps $\eta$ taken by the minimizer. A very small step size indicates that we have not moved far for several steps. Our code will run until the step size becomes smaller than some tolerance value. A smaller value of the tolerance means we get closer to the minimum, however, the computational cost of our minimization is strongly dependent on this tolerance value. We found that a tolerance of the step size of $\eta < 10^{-6}$ produces stable results and manageable low computational cost.

\subsection{Penalty Hyperparameter}
\label{lambda_set}

We will use a k-fold methodology to fit the hyperparameter $\lambda$. K-fold is a well-known method that is standard practice for fitting hyperparameters in sparse regression \citep[e.g.][]{10.5555/2834535}.
The method works by randomly dividing data into $k$ independent subsets of roughly the same size. Then the hyperparameter $\lambda$ is sampled in $k$ independent runs, each time one subset is left out of the minimization and is used to test the model on data it has not seen before. The set left out is called the test set. The rest of the data points are used for running the minimisation algorithm and are referred as the training set. In this work we will use a value of $k=10$, which is standard practice.
Each run explores $\lambda$ with thirty uniformly spread points in $\log_{10} \lambda$ between $\lambda=1$ and $\lambda=N$, to which the case of $\lambda=0$ is added. 

The higher the granularity of $\lambda$ that we explore the more computationally expensive our code becomes. We found by testing that 30 uniformly spread out points in $\log_{10} \lambda$ was enough to find sufficiently smooth curves without a high computational cost. In principle, one could explore larger values of $\lambda$. However in our case models with $\lambda=N$ provided already significantly worse fits than models with smaller $\lambda$ values, which indicates that the penalty was already too large at $\lambda=N$. This is true for all models presented in this paper except for the example of Section~\ref{subsection:toy_model} where we needed to run between $\lambda=0$ and $\lambda=800$.

In order to quantify the quality of fit for a given $\vec C$, we will use the  root mean square error (RMSE) defined as:
\begin{equation}
\label{RMSE}
\text{RMSE}=\sqrt{\frac{\sum_{\alpha=1}^N(F_\alpha(\vec{C},\mathbf{X})-y_\alpha)^2}{N}}
\end{equation}
When $\lambda$ is close to zero, the error in the model of the training set would be small as there is no significant penalty and the model is overfitted. Such a model is poor at predicting results in data that it has never seen before and this translates into a large error on the test set (see Figs.~\ref{lamvsRMSE_test} and~\ref{lamvsRMSE}). For  very large values of the penalty, the model becomes too simple as coefficients are heavily penalised: a model that is too simple will show large error on both the test and the training sets.
When $\lambda$ is large enough to avoid overfitting but not too large that models become too simple, the RMSE of the test set will reach its minimum (as illustrated in Fig.~\ref{lamvsRMSE_test}). If $\lambda_k$ is the value of $\lambda$ where the minimum is for a given k-fold, then $\lambda_\mu=\mu(\lambda_k)$ is an estimate of the optimised value of $\lambda$, where $\mu$ is the {\it mean} operator.

It is common practice in sparse regression to choose a value that is larger than $\lambda_\mu$ by one standard deviation, this is the one-standard-error rule from \cite{10.5555/2834535}. This is done to avoid over fitting due to inaccuracies in the methodology.  In this work we implement a modified version of the one-standard-error rule. Let us define RMSE$_k(\lambda)$ as the RMSE of the $k^{th}$ k-fold as function of $\lambda$.  By construction, RMSE$_k$$(\lambda)$ is minimised for $\lambda=\lambda_k$. If $\sigma(\text{RMSE}_k(\lambda_k))$ is the standard deviation of the collection of RMSE$_k(\lambda_k)$, then for each k-fold we define the optimised value of $\lambda$, $\lambda_{\rm min}$, as:

\begin{equation}
\label{1std_error}
\lambda_{\rm min} = \mu\left(\text{RMSE}_k(\lambda_k)+\sigma(\text{RMSE}_k(\lambda_k))\right)
\end{equation}

In order to find our surviving coefficients, we run the minimization algorithm again on the complete data set, setting $\lambda$ to $\lambda_\text{min}$. With $P$, the number of  coefficients $C_i$ larger than the cutoff value $\mathbf{\nu}$, we define our library of surviving functions $F_S(\mathbf{X})=[F_{S_1}(\mathbf{X}),..,F_{S_P}(\mathbf{X})]$, for which $C_{S_j}>\mathbf{\nu}$ with $1 \le j \le P$. 

The penalty is useful for selecting which coefficients to discard and keep, but once this is done the presence of a penalty term biases all coefficients to smaller values. A penalty rewards smaller coefficients over larger ones, as the size of the penalty increases with the size of the coefficients. Having this in mind, our final model is found by re-running our minimization algorithm using only the functions in $F_S(\mathbf{X})$ and setting $\lambda$ to zero, i.e.\ without penalty. 

\subsection{Example}
\label{subsection:toy_model}

In this section we introduce a simple example to more clearly illustrate our methodology. We build a matrix $\mathbf{X}'$ as in Eq.~\ref{X_matrix}, where each column $\vec{z}_i'$ has thirty points (N=30) and each point is a random number between zero and one. We will use three independent variables, $\vec{z}'_1$, $\vec{z}'_2$ and $\vec{z}'_3$, i.e.\ M=3. We will also build a dependent variable $\vec{y}'$ as:
\begin{equation}
\label{Toy_Model}
\vec{y}'=1.3+2\vec{z}'_1+\mathbf{Noise}
\end{equation}
where the noise comes from a Gaussian distribution centered on zero and with a width of $10\%$  of the standard deviation of $1.3+2\vec{z}'_1$. All of our variables in this example have the same order of magnitude, so there is no explicit need to standardize their units. Hence we use $y'$ and $\vec{z}'$ notation within this section.

Before running the model, we do not know the shape of Eq.~\ref{Toy_Model}. However, let us suppose that we suspect that $\vec{y}'$ should depend on the parameters $\vec{z}'_1$, $\vec{z}'_2$ and $\vec{z}'_3$. As we are uncertain on how to model the dependence between the parameters, we include a large set of functions. In this example, our library of functions includes a constant, linear and quadratic terms only (leaving out cubic terms for simplicity). In total we end up with ten functions (D=10). 

\begin{figure}
\includegraphics[width=\linewidth]{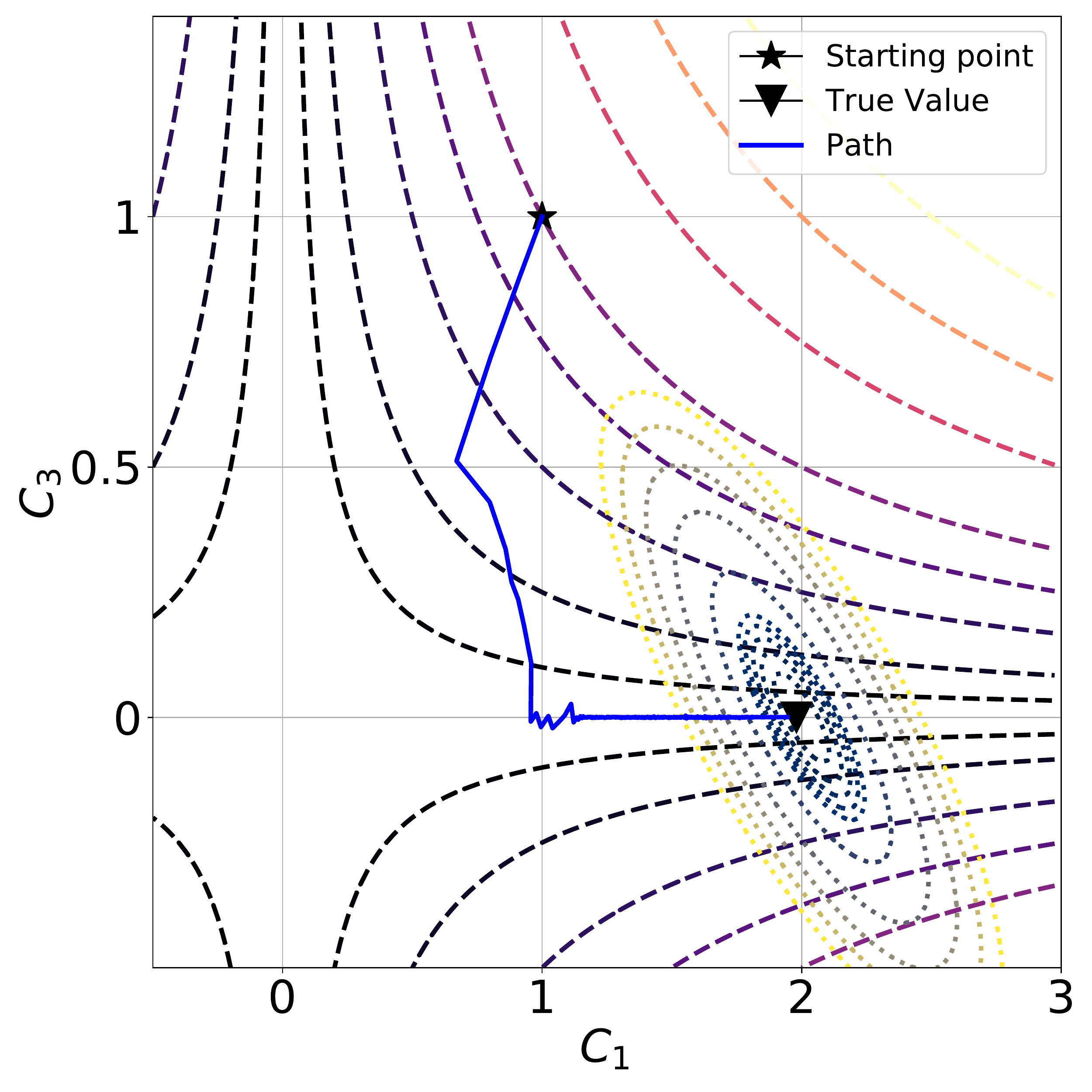}
\caption{Isocontours of the penalty function defined by Eq.~\ref{Penalty} for the two different coefficients $C_1$ and $C_3$ associated with the functions $C_1f^1_1(\vec{z}'_\alpha)=C_1 z'_{\alpha 1}$ and $C_3f^1_3(\vec{z}'_\alpha)=C_3 z'_{\alpha 3}$. The dashed hyperbolic and dotted elliptical lines are the isocontours of our penalty function and of the $\chi^2$ statistic respectively. Given that the gradient is perpendicular to the contour lines, the minimisation routine can efficiently move toward the origin of the plot, and also to one of the axes. Hence the code will quickly reach the minimum if either or both coefficients are zero. }
\label{Contour_plot}
\end{figure}

For the purpose of illustration, we focus on
$C_1$ and $C_3$, the parameters associated with the linear functions of $\vec{z}'_1$ and $\vec{z}'_3$. Fig.~\ref{Contour_plot} follows the trajectory of the minimizer for our example model and for these two coefficients. From Eq.~\ref{Toy_Model} we know that $C_3=0$ and $C_1=2$. The minimizer starts in an arbitrary position (in the case of this example in $C_1=C_3=1$) and follows the trajectory shown by the blue line. The dashed lines represent the contours of both the  $\chi^2$ (elliptic dotted contours) and $P(\vec{C})$ (dashed contours) of Eq.~\ref{Penalty}. A gradient descent algorithm will try to move perpendicularly to these contours, but the modifications in our algorithm allow the path to quickly align to the valley around $C_3=0$. Apparent deviations from this motion come from the fact that we are looking at the 2 dimensional projection of a ten dimensional trajectory.

\begin{figure}
\includegraphics[width=\linewidth]{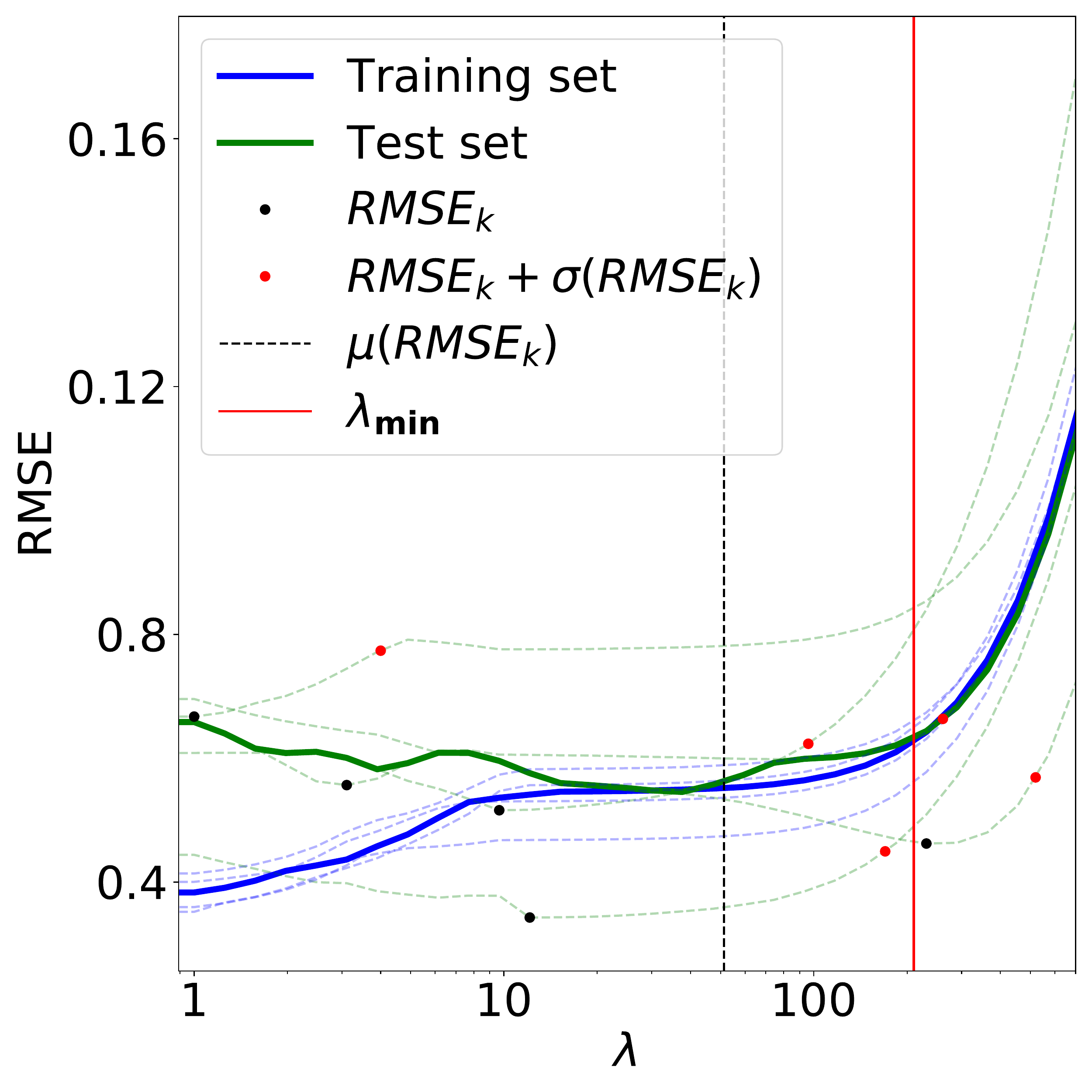}
\caption{Evolution of the RMSE from the best fits of our example model as function of the hyperparameter $\lambda$. The blue and green dashed lines represent the RMSE of the training and test sets respectively. The solid lines represent the median of these curves. When $\lambda$ is close to zero, the training set has a very small error and the test set a comparably larger one: this is due to overfitting of the minimizer and it improves as $\lambda$ grows. The black dots indicate  the minimum RMSE for the test set of each individual k-fold: this is where overfitting was smallest. The black dashed line shows the mean value of the $\lambda$s of the black dots. The red dots are plotted at the values of $\lambda$ given by our modified one-standard-error rule. The red line indicates the mean of $\lambda$s of
these red dots and is our estimate of $\lambda_{\text{min}}$ from Eq.~\ref{1std_error}}.
\label{lamvsRMSE_test}
\end{figure}

Fig.~\ref{lamvsRMSE_test} shows the evolution of the RMSE with respect to $\lambda$ for our example model using the k-fold methodology of Section~\ref{lambda_set}. For this example, we divide the data in to $k=5$ folds.  The blue dashed line correspond to the training set and the green lines to the test set. The solid lines are the median of each set. We explore the hyperparameter $\lambda$ between $\lambda=0$ and $\lambda=800$. This is different to our nominal $\lambda$ range which would be between 0 and N otherwise. This is because the ratio $D/N=10/30$ is much larger in this example than in our nominal set up using the full simulated data set, for which we have hundreds of functions to fit almost 10,000 galaxies (i.e.\ $D/N \sim 0.01$).  This significant change in this ratio of $D/N$ requires a larger penalty to be considered to avoid any overfitting.

When $\lambda$ is close to zero in Fig.~\ref{lamvsRMSE_test}, the RMSE in the training set (blue line) is small, the model is overfitted and therefore bad at predicting the result in data that it has never seen before. This results in the comparably larger error on the test set (green line). For the largest values of the penalty, the model becomes too simple and the error on both the test and the training set begins to increase.
 
Around $\lambda \sim 10$ in Fig.~\ref{lamvsRMSE_test}, the fit of the test set improves and the RMSE reaches its minimum. This is where the model is the least susceptible to overfitting while still capturing the important features of the data set. The black dots indicate the minimum RMSE for the test set of each individual k-fold and the black dashed line shows the mean value of these points, $\mu({\rm RMSE}_k)$. Our optimal value of $\lambda$ is given by $\lambda_{\rm min}$, defined by Eq.~\ref{1std_error} and shown as a vertical red line.
 
\begin{figure}
\includegraphics[width=85mm,height=85mm]{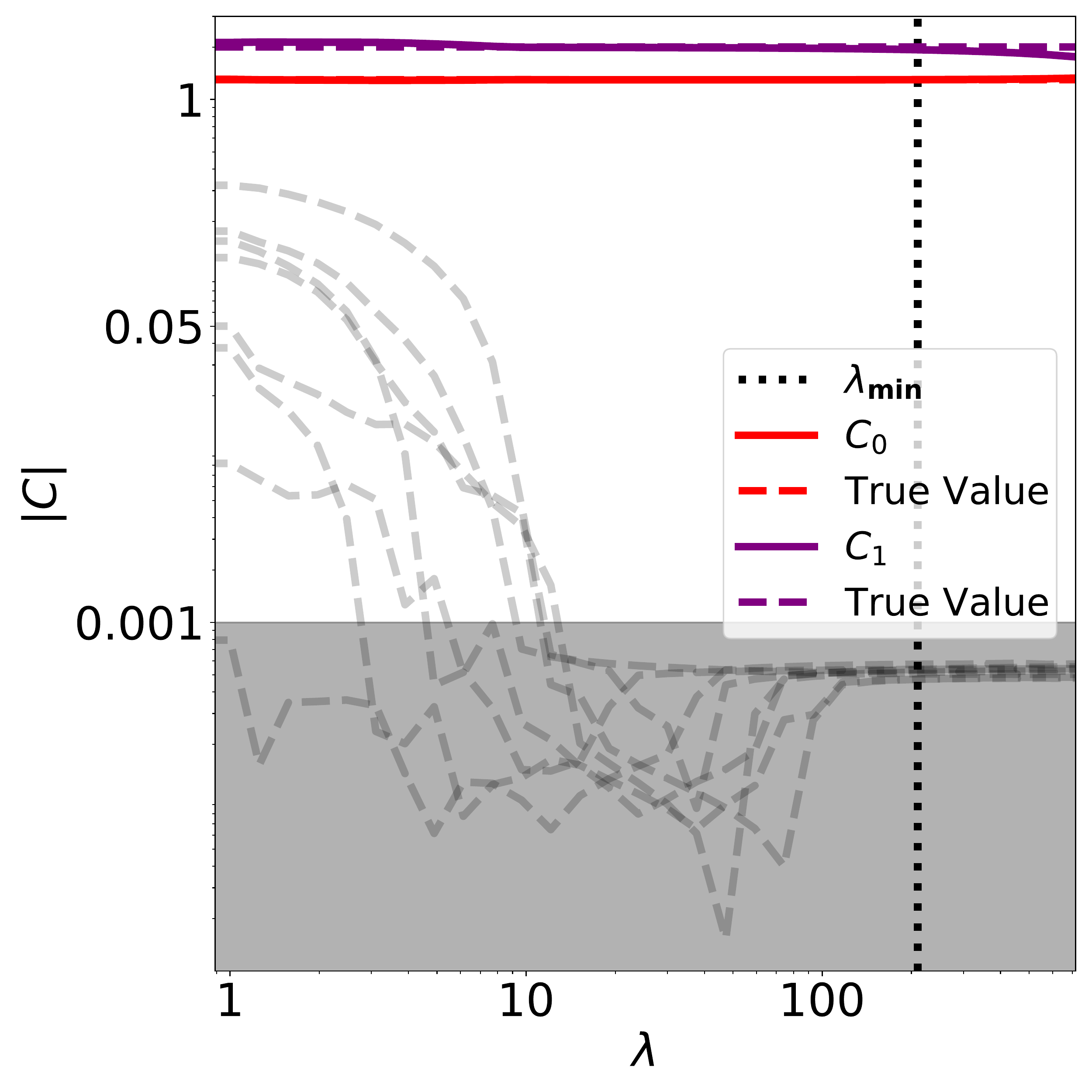}
\caption{Evolution of the absolute value of the best fit coefficient values as function of the hyperparameter $\lambda$. The coloured lines show the value of the mean fit of our independent k-fold runs for our surviving coefficients, i.e.\ those coefficients that are larger than the cutoff value $\nu$ at the optimised value $\lambda_{\text{min}}$ of the hyperparameter $\lambda$. The dashed lines show the true coefficients used to create the data from Eq.~\ref{Toy_Model}. The grey dashed lines show the evolution of the values of the coefficients that were discarded in the final model. The grey shaded area represents our cutoff value $\nu$, below which parameters will be taken out of the fit. The dotted black line represents $\lambda_{\text{min}}$. We note that at $\lambda_{\rm min}$ all coefficients are set to zero except $C_0$ and $C_1$.}
\label{coef_Evo_example}
\end{figure}

Fig.~\ref{coef_Evo_example} shows how the best-fit coefficients of our example model evolve for different values of $\lambda$. Each curve is the mean curve from our 5 different folds. As stated in Section~\ref{sparse_regression}, the code will not set parameters exactly to zero but to a very small value which is determined by the parameter $\epsilon$ of Eq.~\ref{Penalty}. Fig.~\ref{coef_Evo_example} shows that in this example the value is $\sim 6 \times 10^{-4}$ (this is true for both the example model and the galaxy data set as it only depends on $\epsilon$). Therefore we select a cutoff value of $\nu=1\times10^{-3}$. This is is shown as the grey shaded region in Fig.~\ref{coef_Evo_example}.
 
The coloured lines correspond to the coefficients that were above the cutoff value $\nu$ at $\lambda_{\rm min}$, and therefore included in the final model. At $\lambda=0$, all coefficients are above the cutoff threshold due to over-fitting. As $\lambda$ grows, coefficients drops below the threshold value and at $\lambda_{\rm min}$ all coefficients other than $C_0$ and $C_1$ have been discarded. This is expected as $C_0$ and $C_1$ are the non-zero coefficients used in building $\vec{y}'$ according to Eq.~\ref{Toy_Model}. The grey dashed lines are the coefficients that were below $\nu$ at $\lambda_{\rm min}$ and therefore discarded. The vertical dashed line corresponds to the optimized value $\lambda_\text{min}$.

The final model selected by the algorithm is:
\begin{equation}
\label{Toy_Model_Final}
\vec{F}(\vec{C},\mathbf{X}')=1.27+1.98 \, \vec{z}'_1
\end{equation}
Considering that this a fit to data generated using Eq.~\ref{Toy_Model} with 10\% Gaussian distributed noise,

we can conclude that our algorithm generated a sparse and accurate representation of the data.

\section{Data Set}
\label{DataSet}
This section introduces the data set that is used in our analysis. In \S\ref{intro_EAGLE}, we introduce the hydrodynamical simulation, which data is used to train our model on. In \S\ref{Data_selection}, we describe the selection of the galaxies considered, as well as the method used to address inconsistencies in the classification of galaxies between different snapshots of the simulation. \S\ref{Input_param} and \S\ref{specific angular momentum} introduce the set of variables that form $\vec{x}'_\alpha$  of Eq.~\ref{X_matrix}: the former introduces all variables associated with the mass of the host halo and the latter with its angular momentum. Finally \S\ref{Experiment_list} 
lists the four models considered, which differs from one another by the set of variables used to define $\vec{x}'_\alpha$.

\subsection{The EAGLE simulation}
\label{intro_EAGLE}
Hydrodynamical simulations provide powerful insight into the galaxy formation process. The resulting database catalogues DM halos and their connection to baryonic properties such as stellar mass. In this paper, we use the Evolution and Assembly of Galaxies and their Environments \citep[ EAGLE,][]{2015MNRAS.446..521S,2015MNRAS.450.1937C} simulations, a suite of hydrodynamical simulations built inside cubic periodic volumes. We use the largest of these volumes, corresponding to a box of 100 comoving Mpc of length.

The simulation runs using a modification of the GADGET 3 code described in \cite{2005MNRAS.364.1105S}. The code uses Smooth Particle Hydrodynamics methods to model the mechanics of the baryon fluid. In order to compute the gravitational potential, the code uses a combination of a Particle Mesh (at large scales) and a hierarchical Tree algorithm (at grid and subgrid scales). The details on the modifications can be found in \cite{10.1093/mnras/stv2169}. The simulations are built using the Planck cosmology \citep{2014A&A...571A...1P}.

Baryonic physical processes that cannot be solved directly are implemented into the simulation as sources and sink terms, where energy and matter are either absorbed or injected locally into the simulation. These subgrid models should depend only on the local property of the gas. The subgrid models implemented account for radiative cooling \citep{2009MNRAS.393...99W}, star formation \citep{2008MNRAS.383.1210S}, star formation feedback \citep{2012MNRAS.426..140D}, black hole growth \citep{2015MNRAS.454.1038R,2005Natur.435..629S},  Active Galactic Nuclei feedback \citep{2009MNRAS.398...53B} and chemical enrichment \citep{2009MNRAS.399..574W}.
The uncertain parameters of the subgrid models need to be calibrated, which is done by choosing the values that would reproduce the galaxy mass function at $z$=0.1, the galaxy size-stellar mass relation and the black hole mass-stellar mass regression. Discussion of the calibration process can be found in \cite{2015MNRAS.450.1937C}.

Haloes are defined using a Friends-of-Friends algorithm \citep[FoF; e.g.][]{1984MNRAS.206..529E} with a linking length of $b$=0.2, i.e.\ all particles that can be linked together with an inter-particle distance smaller than 0.2 times the mean inter-particle distance form a halo. Once the halos have been identified, the SUBFIND algorithm \citep{2001MNRAS.328..726S} identifies the self-bound local overdensities of each FoF group as subhalos. The subhalo that contains the particle with the lowest value of the potential energy will be defined as the central sub-halo.

The simulation information is saved at 29 redshifts from $z$=20 to $z$=0 (i.e.\ 29 snapshots), and is used to build merger trees, which connect a halo to its progenitors at earlier redshifts \citep{2017MNRAS.464.1659Q}. The main progenitor of a halo is defined as the progenitor with the largest mass at all earlier outputs. We use these main progenitors to track the mass evolution of a DM halo (Section~\ref{Input_param}). 
Note that when two halos pass close to each other without merging they could momentarily belong to the same FoF group. As a consequence, the mass and the subhalo chosen as the central may be inconsistent at this snapshot when compared to the one immediately before or after the interaction \citep{2015MNRAS.454.3020B}. We introduce a scheme to clean such issues from the input data in section~\ref{Data_selection}.

\subsection{Data selection}
\label{Data_selection}

Our data set consists of central galaxies inside halos with a mass larger than $M_{200,C} > 10^{11.1}$M$_\odot$.
$M_{200,C}$ corresponds to the mass inside the radius $R_{200,C}$ of a halo, which is the radius within which the density is 200 times the critical density of the universe. The stellar mass of a galaxy is measured as the baryonic mass contained inside a sphere of 30 proper kpc around the centre of potential of the halo.

Baryonic processes inside halos can affect their measured DM properties \citep{2013MNRAS.429.3316B,10.1093/mnras/stv1341}.  If we run our code using the properties of the DM found in a hydrodynamical simulation, we risk including biases by fitting the stellar mass using a property that has been modified by the presence of baryons (this modification would be correlated with the stellar mass as the halos with more baryons would be more modified). To avoid this bias, it is common practice to extract all DM input properties from a DM only simulation generated with the same initial conditions, box size and resolution as the full hydrodynamic simulation.

The matching between the hydrodynamic and DM-only simulations is described in \cite{2015MNRAS.451.1247S}. The 50 most bounded DM particles of each halo in the hydrodynamic simulation are found. If a halo in the DM-only simulation has at least half of those most bound particles it is considered its analogue. Using this method, $99\%$ of the halos with $M_{200,C}>1\ \times 10^{11.1} $M$_{\odot}$ are matched. 
We collect information about the host DM halo at different redshifts (Section~\ref{Input_param})and require that our halos are present in all snapshots. With this in mind, we only use galaxies with a progenitor defined at $z$=4 . Our full sample consists of  9521 galaxies.

Inconsistencies between snapshots are a well-known characteristic of the merger trees \citep{2015MNRAS.454.3020B} created by running the halo finder separately on each snapshot. When two halos interact some of the particles of one can be assigned to the other regardless of where they belonged in past snapshots. One consequence is that small central halos can be considered satellites of a larger halo if they are close to each other at a given snapshot. In EAGLE, $M_{200,C}$ is only computed for central halos, which means that they will not have a value of $M_{200,C}$ at these snapshots. 

When this happens we interpolate the value of $M_{200,C}$ in the missing slices using the following methodology: we look for the $M_{200,C}$ value of both the nearest earlier and later redshifts where the halo was still central. We use these values to do a linear interpolation of $M_{200,C}$ in the missing slice. The nearest earlier redshift is always well defined (as at z=0 all of our selected subhalos are central); however, a small subset of galaxies have a non-central progenitor at their largest redshifts, and therefore their nearest latter subhalo is not necessarily well defined. In these cases we select the third to last and second to last halos and perform our interpolation with those. We follow a similar procedure to correct the angular momentum of halos that are not considered central in a given slice. We found that the value of the angular momentum can have drastic variations when compared to its value at the surrounding redshift slices, which is due to the number of particles assigned to the halo changing significantly when it is misclassified as a subhalo.

The black lines of Fig.~\ref{DM_input} shows the halo mass history relative to the halo mass at redshift zero of four halos from $z=4$ and to $z=0$. The figure shows that different halos have very different formation histories. We will explore whether galaxies that have followed different halo formation paths will end up having different residuals in the SMHM relation.

\subsection{DM Mass}
\label{Input_param}

Once we have selected the galaxies in our data set, we define the $M$ parameters of the DM halo that are used to build the matrix $X'$ of Eq.~\ref{X_matrix}.
The first variable accounted for is the halo mass at redshift zero (or any variable highly correlated with it), as the SMHM relation explains most of the scatter in the stellar mass. We will denote the Halo Mass input variable of a galaxy as $M_0'$ and define it as

\begin{equation}
\label{M0}
M_0'=\log_{10}(M_{200}^c(z=0) /M_\odot).    
\end{equation}
We use Eq.~\ref{Normalisation} to standardize the units and denote the halo mass in standardized units as $M_0$.

There is significant scatter around the SMHM relation due to their varied formation history, therefore we should also add parameters that are good estimators of the mass evolution of the DM host halo. This can be done by adding the halo mass of the main progenitor of a host halo at different redshift slices into our $X'$ matrix. 

The EAGLE simulation has 19 snapshots between z=0 and z=4 \citep{2016A&C....15...72M}. However, information between redshift slices that are close to each other is strongly correlated as halos have not evolved significantly. Keeping this in mind and given that the computational cost of running the minimizer increases exponentially with the number of parameters, we only use a subset of the available redshifts. The ten redshifts slices that we use as inputs are $z_\mathbf{in}$=[0.0,~0.18,~0.37,~0.62,~1.0,~1.26,~1.74,~2.48,~3.02,~3.98].

Sparse regression methods work best if variables are independent, therefore we will use the ratio between the mass at a given redshift and the mass at redshift zero (so that the significant correlation of the mass at a given redshift and its mass at $z=0$ is removed). We will denote these variables as $(M_z/M_0)'$ and they are defined as:
\begin{equation}
\label{ratio_z}
(M_z/M_0)'=\log_{10} \left( {\frac{M_{200}^c(z)}{M_{200}^c(z=0)}} \right)
\end{equation}
We then use Eq.~\ref{Normalisation} to standardise the units and form $M_z/M_0$. 

\begin{figure}
\includegraphics[width=\linewidth]{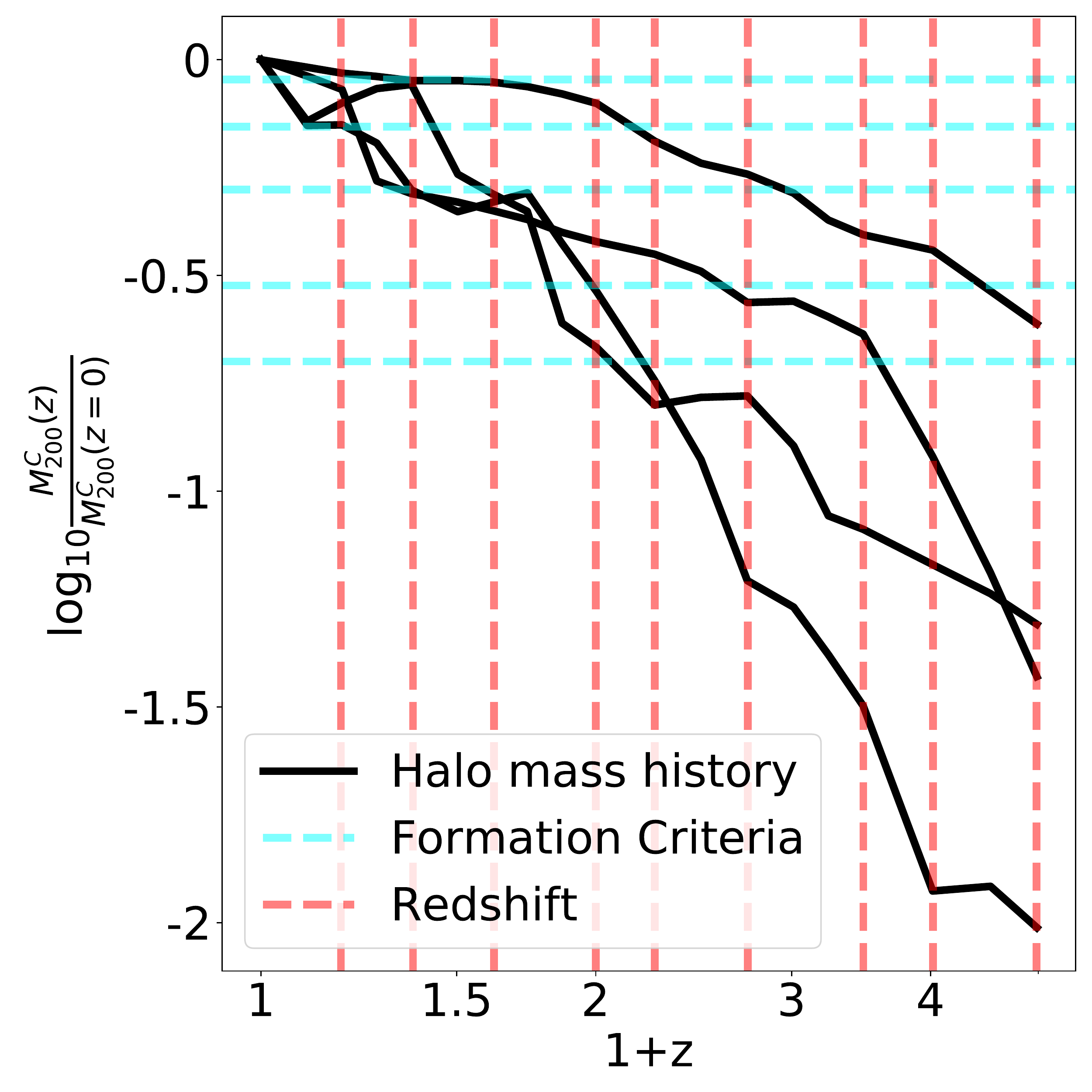}
\caption{Halo mass history of four halos between $z=4$ and $z=0$ (black lines), as given by the ratio of the mass at $z$ to its present day value. The vertical red dashed lines indicate the redshifts used in the analysis (i.e.\ $z_\mathbf{in}$). The $(M_z/M_0)'$ parameters are given by the intersection of the red and black lines. The blue horizontal lines correspond to a constant mass ratio of 90\%, 70\%, 50\%, 30\% and 20\% (from top to bottom). The formation criterion parameters $\mathbf{FC}'_p$ can be visualized as the intersection between the blue and the black lines.}
\label{DM_input}
\end{figure}

An alternative approach to characterise halo evolution is the formation time \citep{1994MNRAS.271..676L}, defined as the time at which a halo has assembled half of its present day mass. We generalize this idea to define five formation criteria ($\mathbf{FC}'$) by finding the redshifts (instead of times) at which the DM halo has assembled $20\%$, $30\%$, $50\%$, $70\%$ and $90\%$ of its mass respectively. The set of all five formation criteria for our sample will be referred to as $\mathbf{FC}'_p$, where $p$ denotes the percentage used. Fig.~\ref{DM_input} shows a set of horizontal blue lines corresponding to halo mass ratios (Eq.~\ref{ratio_z}) of 90\%, 70\%, 50\%, 30\% and 20\%. The redshifts at which each formation history curve (black solid line) intersects this blue horizontal lines is a visual representation of $\mathbf{FC}'_p$. The redshifts that correspond to a given formation criteria are found by performing a linear interpolation of the halo mass ratios. As with all parameters, a final step is to standardise the units using Eq.~\ref{Normalisation} and to form $\mathbf{FC}_p$.

\subsection{Specific angular momentum}
\label{specific angular momentum}

There is a well known observational scaling relation between the angular momentum of a galaxy and its stellar mass \citep{Fall_2013}. It is a matter of discussion, however, how much of a role the angular momentum history of a dark matter halo plays in determining the specific angular momentum of its host galaxy. \cite{10.1093/mnras/stw1286} finds strong correlations between both parameters using the EAGLE simulation. However, \cite{Danovich_2015} suggest that the specific angular momentum of gas and dark matter undergo different formation histories, which would suggest that any correlation between them is a by-product of a third correlation with other parameters like the mass formation history. 
Having this in mind, we will generate candidate models that also include specific angular momentum input parameters on top of the mass evolution parameters. 

Angular momentum evolution is included in our methodology by computing the halo specific angular momentum vector, {\bf $\vec{\mathbf{j}}$}, defined within a radius $R$ and for each redshift slice $z$ as:
\begin{equation}
\label{sAM from N-body}
\vec{\mathbf{j}}(R,z)=\frac{\sum_i m_i(\vec r_i-\vec r_{c})\times(\vec v_i-\vec v_{c})}{\sum_i m_i},
\end{equation}
where $\vec{r}_i$ and $\vec{v}_i$ are the position and velocity vectors of each particle within a radius $R$ of the centre of mass, $m_i$ is the mass of the particle, and $\vec{r}_c$, $\vec{v}_c$ are the position and velocity of the centre of mass of the halo. We will use different values of $R$ in order to capture the angular momentum evolution of the full halo and of its centre separately. The values of $R$ that are included in our model are $R_{200}^C$,  $\frac{R_{200}^C}{2}$ and $\frac{R_{200}^C}{5}$, which are all functions of redshift.

The specific angular momentum defined in Eq.~\ref{sAM from N-body} correlates strongly with the mass of the halo: this is driven by the scaling relations $|\vec r| \propto M^{1/3}$ and $|\vec v| \sim \sqrt{\frac{GM}{R}}\propto M^{1/3}$. To avoid  strongly correlated variables in our parameter set, we define the following specific angular momentum parameter:
\begin{equation}
(S(R,z))'=\log_{10}(|\vec{\mathbf{j}}(R,z)|)-\frac{2}{3} \log_{10}(M^C_{200}(z)/M_\odot)  \,
\end{equation}
where $|\vec{\mathbf{j}}(R,z)|$ is the norm of $\vec{\mathbf{j}}(R,z)$.

Given that the angular momentum is a vector, we need two types of variables to describe it: one capturing its magnitude and the other one its direction. Therefore, we will also include the change in the angle, $\Theta$', defined as the scalar product between the halo specific angular momentum at redshift $z$ w.r.t.\ the one at the present time, i.e.:
\begin{equation}
\label{apin_direction}
(\Theta(z))'=\frac{\vec{\mathbf{j}}(R_{200}^C,z) \cdot \vec{\mathbf{j}}(R_{200}^C,0)}{|\vec{\mathbf{j}}(R_{200}^C,z)||\vec{\mathbf{j}}(R_{200}^C,0)|}
\end{equation}
Note that by definition $(\Theta(z=0))'=1$ for all galaxies and hence we only include $(\Theta(z>0))$ in our list of variables. As with all other variables we use Eq.~\ref{Normalisation} to standardize the units and form the scalars $S(R,z)$ and $\Theta(z)$. We form the following library of {\bf $\vec{\mathbf{j}}$} parameters for each halo $i$ at each redshift:
$S_i(R_{200}^C,z)$, $S_i(\frac{R_{200}^C}{2},z)$, $S_i(\frac{R_{200}^C}{5},z)$ and $\Theta_i(z)$.

The evolution of our specific angular momentum parameters has significant statistical noise and so it is smoothed across different redshifts using a Gaussian Kernel.

\subsection{Models}
\label{Experiment_list}
The four models considered in this work are:

\begin{enumerate}

    \item {\bf Mass ratio}: This model includes values of the halo mass at redshift zero, $M_0$  (Eq.~\ref{M0}), and the halo mass ratios, $M_z/M_0$ (Eq.~\ref{ratio_z}), that parameterise the DM halo mass evolution. With 10 different redshift slices, this gives a total of $M=10$ input parameters, resulting in a total of D=286 functions to minimize over (Eq.~\ref{N_function}). 
    \item {\bf Formation criterion}: In this model, the DM ratios are replaced by the formation criterion $\mathbf{FC}_p$, defined in Section~\ref{Input_param}. This model uses, as parameters, 5 values of $\mathbf{FC}_p$ (with p=[90,~70,~50,~30,~20]) and the halo mass at redshift zero, $M_0$, resulting in $M=6$ and $D=84$ functions to minimize over.
    \item {\bf Mass ratio and {\bf $\vec{\mathbf{j}}$}}: In this model we add the specific angular momentum parameters {\bf $\vec{\mathbf{j}}$} (and more specifically $S(R_{200}^C,z)$, $S(R_{200}^C/2,z)$, $S(R_{200}^C/5,z)$ and $\Theta(z)$ at each of the ten snapshot considered), to the library of free parameters of the mass ratio model. The library of functions contains the linear, quadratic and cubic terms of the Halo mass evolution parameters $M_z/M_0$. Only the linear terms of the specific angular momentum parameters are included. To include all the quadratic and cubic terms would result in $D=23426$ functions to minimize over, which at the moment is too computationally expensive for our algorithm. Hence we will only include linear terms for the specific angular momentum parameters, ending up with a total of $D=326$ functions to minimize over.
    \item {\bf Formation criterion and {\bf $\vec{\mathbf{j}}$}}: This model is similar in spirit to model (iii), but we add the terms of the specific angular momentum parameters, $\vec{\mathbf{j}}$, to the library of free parameters of the formation criterion model instead. As with model (iii), we consider only the linear terms of the specific angular momentum parameters, ending up with $D=123$ functions to minimize over. 
   
\end{enumerate}

\section{Running the Algorithm}
\label{Runing_alghoritm}
In this section we present some specific aspects of applying the methodology presented in Section~\ref{section:Sparse_Regresion} to the data described in Section~\ref{DataSet}. In particular tests of the consistency of the algorithm are considered: we evaluate the impact of the chosen $\epsilon$ parameter in Section~\ref{lambda set} and discuss the uncertainty of the parameter models in Section~\ref{sec:uncertainty}. The model results are presented and discussed in Section~\ref{results}.

\subsection{Training, holdout and test sets}
\label{Preparing}

The data is randomly divided into two, the training set and the holdout set. The training set contains 85$\%$ of the data and is used by the algorithm to build the model. The remaining 15$\%$  constitute the holdout set and is not used until the model is completed. The final model is applied to the holdout set to test its accuracy by considering data not used in the building of the model and therefore is unbiased to over-fitting.

Note that the holdout data set is different from the test sets used for estimating the optimal value of the hyperparameter $\lambda$ in the k-fold methodology of Section~\ref{lambda_set}. The latter constitutes sets drawn from the training set that are systematically kept out of the minimizations done while exploring the $\lambda$ parameter space and are used to determine $\lambda_{min}$. They are part of the methodology for building our model. The holdout set, on the other hand, is kept out of this methodology completely and is used to evaluate the final model once it is built.

\subsection{Penalty Hyperparameter}
\label{lambda set}

This section applies the methodology used for optimizing the hyperparameter $\lambda$, as introduced in Section~\ref{lambda_set}. It discusses the impact of the assumed value for the parameter $\epsilon$, used in the penalty function (Eq.~\ref{Penalty}). 
 
\begin{figure}
\includegraphics[width=\linewidth]{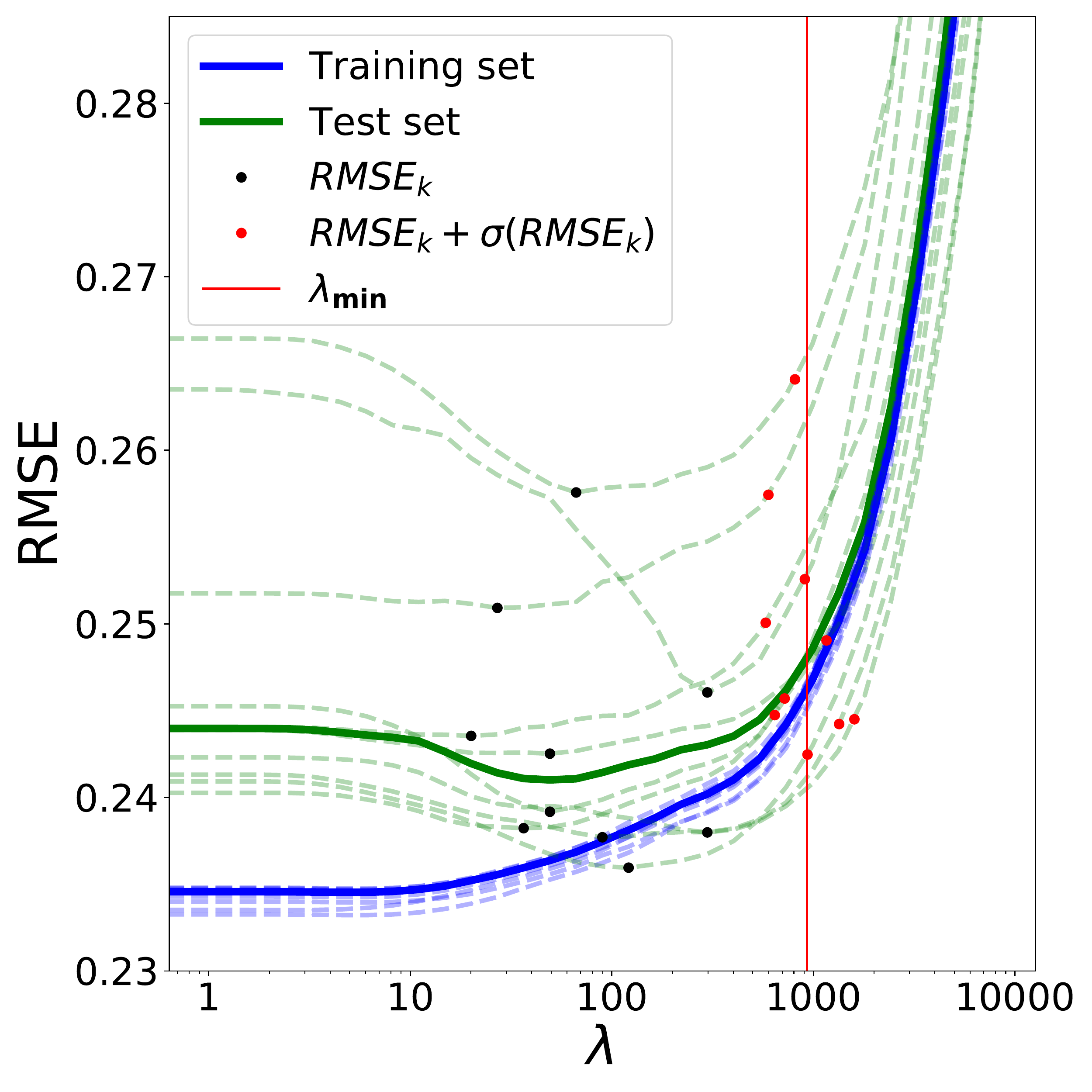}
\caption{Evolution of the RMSE (Eq.~\ref{RMSE}) of the mass ratio model (\S\ref{Experiment_list}) as a function of hyperparameter $\lambda$ for our nominal EAGLE data set (Section~\ref{DataSet}). The blue and green dashed lines represent the training and test sets respectively. The solid lines represent the median of these curves. The black
dots show the minimum of the dashed lines ($\text{RMSE}_k$) and the red dots the one-standard-error rule correction from Eq.~\ref{1std_error}. The red solid line corresponds to the mean $\lambda$ of the red dots and is our estimate of $\lambda_\text{min}$. }
\label{lamvsRMSE}
\end{figure}

From Section~\ref{lambda_set}, the optimal value of $\lambda$, $\lambda_{min}$, is determined using a k-fold method with $k=10$ folds. Each fold runs independently and in parallel on different computer nodes. Fig.~\ref{lamvsRMSE} 

shows the evolution of the RMSE of the mass ratio model (\S\ref{Experiment_list}) as function of the hyperparameter $\lambda$. The green and blue dashed lines correspond to the test and training sets respectively. Each test set (green dashed lines) has around 800 points, which is around $10\%$ of our training data set,  and the minimization runs with D=286 free parameters. 
The green dashed lines show some spread in their amplitudes, which are correlated with their value at $\lambda=0$. This spread is a consequence of dividing the subsets randomly. Some subsets will contain a larger amount of points that are well predicted by the model and will, therefore, have smaller errors. 

As we saw with Fig.~\ref{lamvsRMSE_test}, the RMSE of the training set is smaller when $\lambda \sim 0$ as overfitting makes the model agree unreasonably well with the data it uses for the fitting. In contrast when the model is tested on data it has not seen before, the RMSE is larger, as shown by the comparatively larger error on the test set. As $\lambda$ increases, the error on each test set decreases and eventually, reaches a minimum ($\text{RMSE}_k$) around $\lambda \sim 100$, as shown by the black dots in Fig.~\ref{lamvsRMSE}. This is where the model is least susceptible to overfitting, while still capturing the important features of the data set.

The red dots in Fig.~\ref{lamvsRMSE} show the correction obtained with the one-standard-error rule from Eq.~\ref{1std_error}. The plot shows that these points are to the right of the minimum value of the green dashed lines, however the differences in RMSE between the actual minima and the red dots are small. This means that the resulting models are simpler (and therefore more explainable) and with comparable accuracy. The red solid line is the optimized value of the hyperparameter, $\lambda_\text{min}$, as estimated using Eq.~\ref{1std_error}. 

\begin{figure}
\includegraphics[width=85mm,height=85mm]{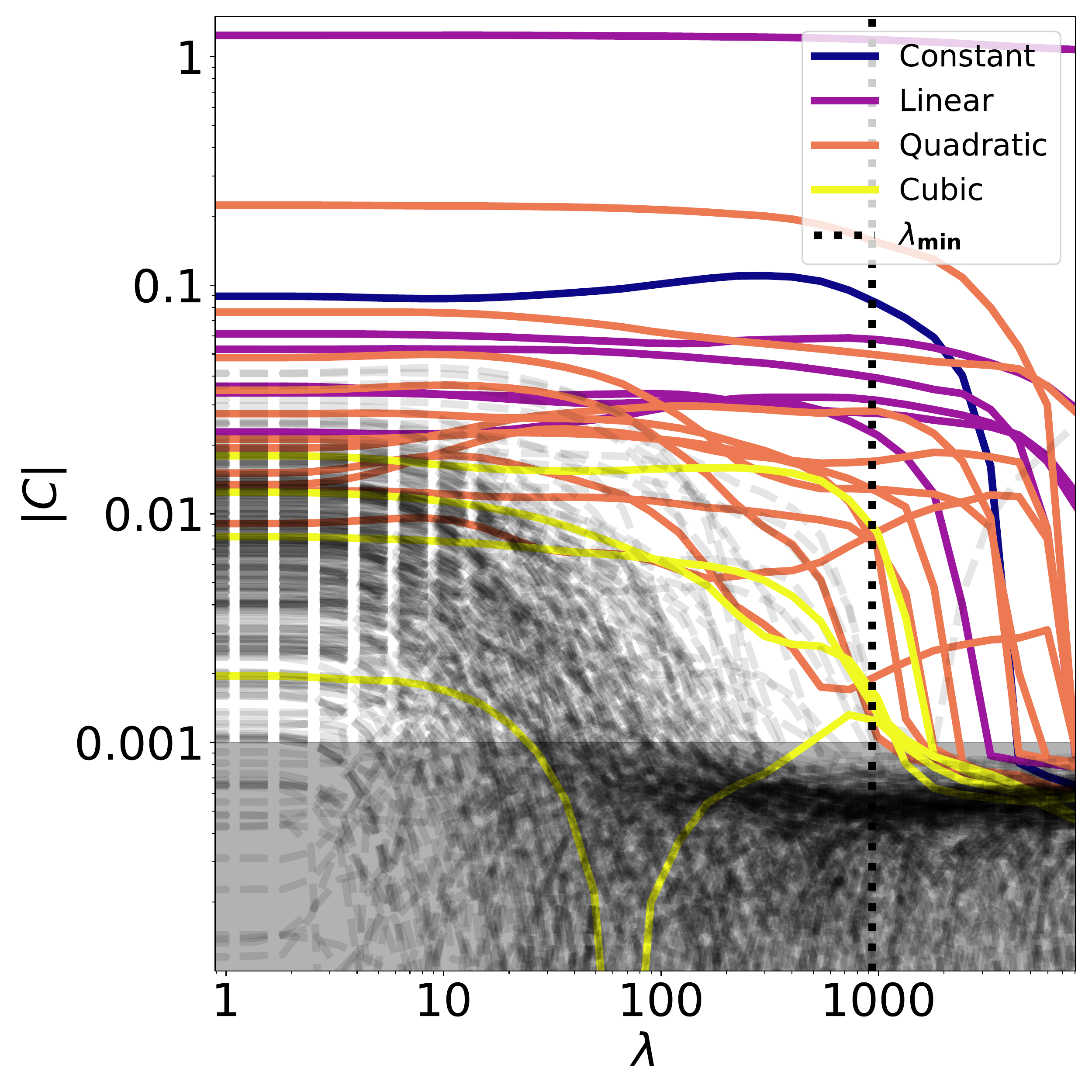}

\caption{Evolution of the best fit value of each coefficient of the mass ratio model as a function of the hyperparameter $\lambda$. The coloured lines show the values of the accepted coefficients and the black dashed lines represent the rejected coefficients. The vertical dotted black line shows the value of $\lambda_\text{min}$ and the grey shaded area represents the region bellow the cutoff value $\nu$: all coefficients above the shaded region at $\lambda_\text{min}$ are retained by the model and represented by coloured lines. 
}
\label{coef_Evo}
\end{figure}

Fig.~\ref{coef_Evo}

shows the evolution of the coefficients $C_i$ of Eq.~\ref{matix_definition} of the mass ratio model (\S\ref{Experiment_list} as a function of the hyperparameter $\lambda$.
The vertical black dotted line shows the value of $\lambda_{min}$ found by our algorithm. Each coloured line correspond to a coefficient that is above the cutoff value $\nu$ at $\lambda_\text{min}$, with $\nu$ represented as the boundary between the white and grey regions of the plot. The grey dashed lines correspond to the coefficients that are below $\nu$ at $\lambda_\text{min}$ and therefore discarded. The figure shows that coefficients that have been discarded have a value of around $0.0005$ or lower (shown by the average value of the black dashed lines at large values of $\lambda$), given that our cutoff value is $0.001$ there is a distinct separation between the chosen coefficients and those discarded. 

Different $C_i$ coefficients are fitted by the minimizer with different orders of magnitude\footnote{As our input variables are not Gaussian, several parameter values are above one standard deviation. In a standardized space this will mean that they will be larger than one. As a consequence linear, quadratic and cubic coefficients will require different scales to make similar contributions.}. Therefore we need to make sure that the value of the parameter $\epsilon$ of the penalty term (Eq.~\ref{Penalty}), which determines how close to zero unnecessary parameters get in the minimization, is such that discarded coefficients are well below the cutoff value $\nu$, and are close enough to zero that they can be separated from useful coefficients. 

Nominally we use a value of $\epsilon=10^{-3}$, which, as shown in Fig.~\ref{coef_Evo}, corresponds to the minimizer setting unused  parameters to a value as small as $\approx 6\times 10^{-4}$. This is comparable to the findings of the example presented in Section~\ref{subsection:toy_model}.
A very small value of $\epsilon$ increases the computational time significantly given that parameters need to be driven further toward zero. Our choice represents a value of $\epsilon$ that is small enough to get parameters close enough to zero while not being so small that the code becomes too expensive to run.

\begin{figure}
\includegraphics[width=85mm,height=85mm]{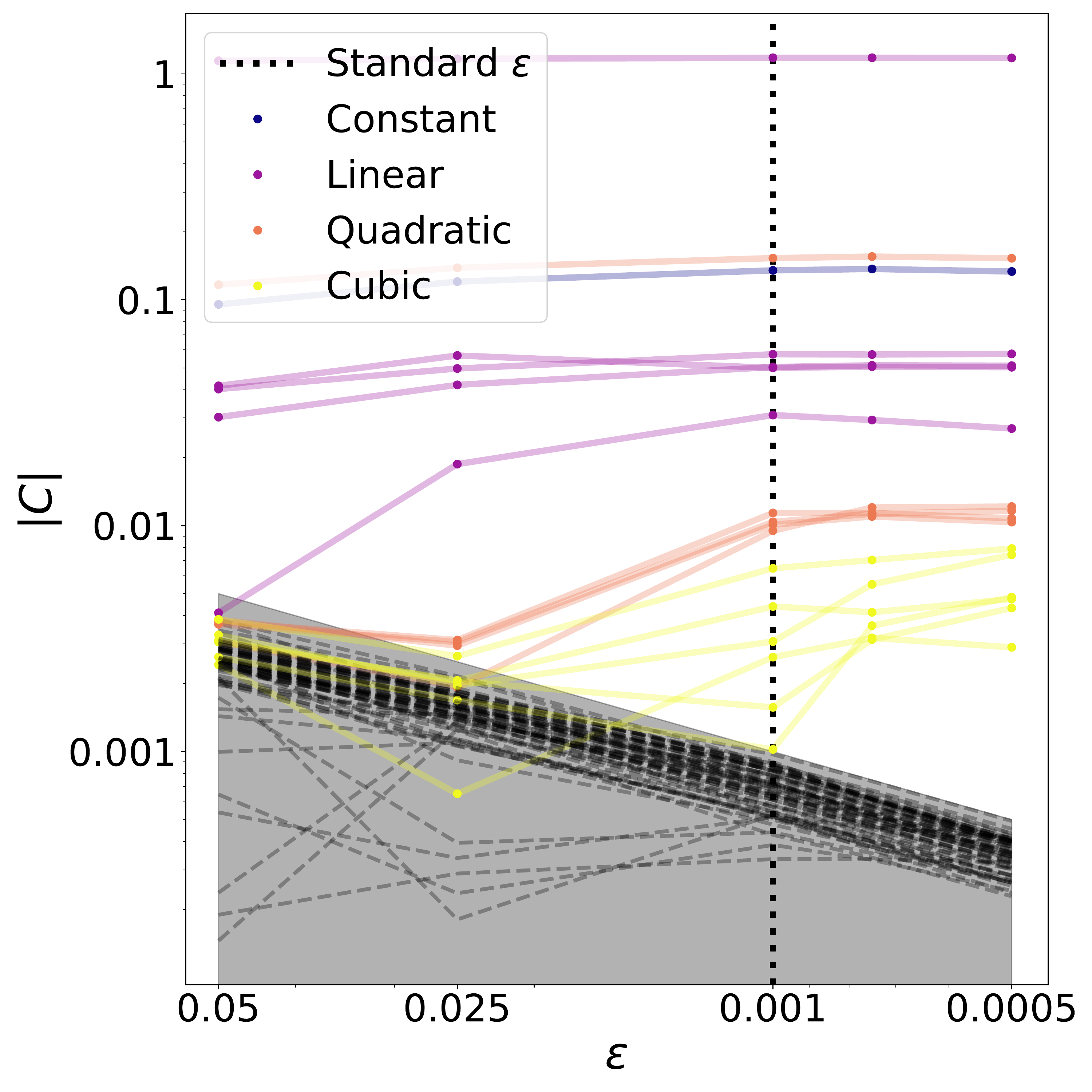}
\caption{
Best fit coefficients for the formation criterion model for five different values of the $\epsilon$ parameter (Eq.~\ref{Minimisation}).
This parameter determines how close to zero coefficients get before their contribution to the penalty is negligible. The cutoff value $\nu$ is set as $\nu=\epsilon$ for each run. The black dotted line shows the value of $\epsilon$ used in our standard configuration. When $\epsilon$ is large, all coefficients are above the cutoff value $\nu$. For $\epsilon=5 \times 10^{-4}$, all kept coefficients are significantly larger than $10^{-3}$, indicating the adequacy of our nominal choice for $\epsilon$.}
\label{epsilon_evo}
\end{figure}

To test what impact the value chosen for $\epsilon$ has, we consider the formation criterion model (\S\ref{Experiment_list}). This model has less free parameters than the mass ratio model ($D=84$ versus $D=286$) and hence requires significantly less computational time, enabling an adequate $\epsilon$ parameter space to be explored.
Fig.~\ref{epsilon_evo} shows the resulting coefficients after running our full algorithm using five different values of $\epsilon$ using the formation criterion model. 

The coloured lines show the parameters that are above the cutoff value $\nu$ in the model built with $\epsilon=10^{-3}$ (our standard value) and represent the variables that where chosen by the algorithm. The grey dashed lines correspond to the values rejected at $\epsilon=10^{-3}$. The cutoff value $\nu$ depends on how close parameters get to zero and therefore it is a function of $\epsilon$. For the propose of illustration, we set $\nu=\epsilon$. 
 
For larger values of $\epsilon$, there is no clear cut between discarded coefficients and most of the cubic and quadratic terms end up in our model. On the opposite end at $\epsilon=5 \times 10^{-4}$, all accepted coefficients are significantly greater than the cutoff value.

While the difference between useful and useless coefficients is clearer at $\epsilon=5 \times 10^{-4}$, our standard configuration with $\epsilon=10^{-3}$ seems to work just as well while being significantly faster to run.

\subsection{Uncertainty on the models}
\label{sec:uncertainty}

Several of our coefficients $C_i$ are associated with functions of the same form but with inputs from different redshifts (see Section~\ref{Input_param}). If the halo mass does not vary significantly between adjacent redshift slices, then the corresponding polynomial functions $f_i(x)$ are likely to show some correlation between them.

In general different order combinations of correlated terms will also be correlated. Considering the above statements, it is possible that the parameter space of $C_i$ coefficients has several local minima. This could be an issue for gradient decent algorithms, as by construction they will converge only toward the closest minimum. In practice we are satisfied with any reasonable minimum: for example, we do not have a preference between a feature being explained by the halo mass ratio at one specific redshift versus that of an adjacent redshift slice. 

This, however, means that there might be slight variations in the surviving parameters of different models depending on the starting point of the minimization and depending on the specific selection of the training set. We test for both aspects in turn.

\begin{figure}
\includegraphics[width=\linewidth]{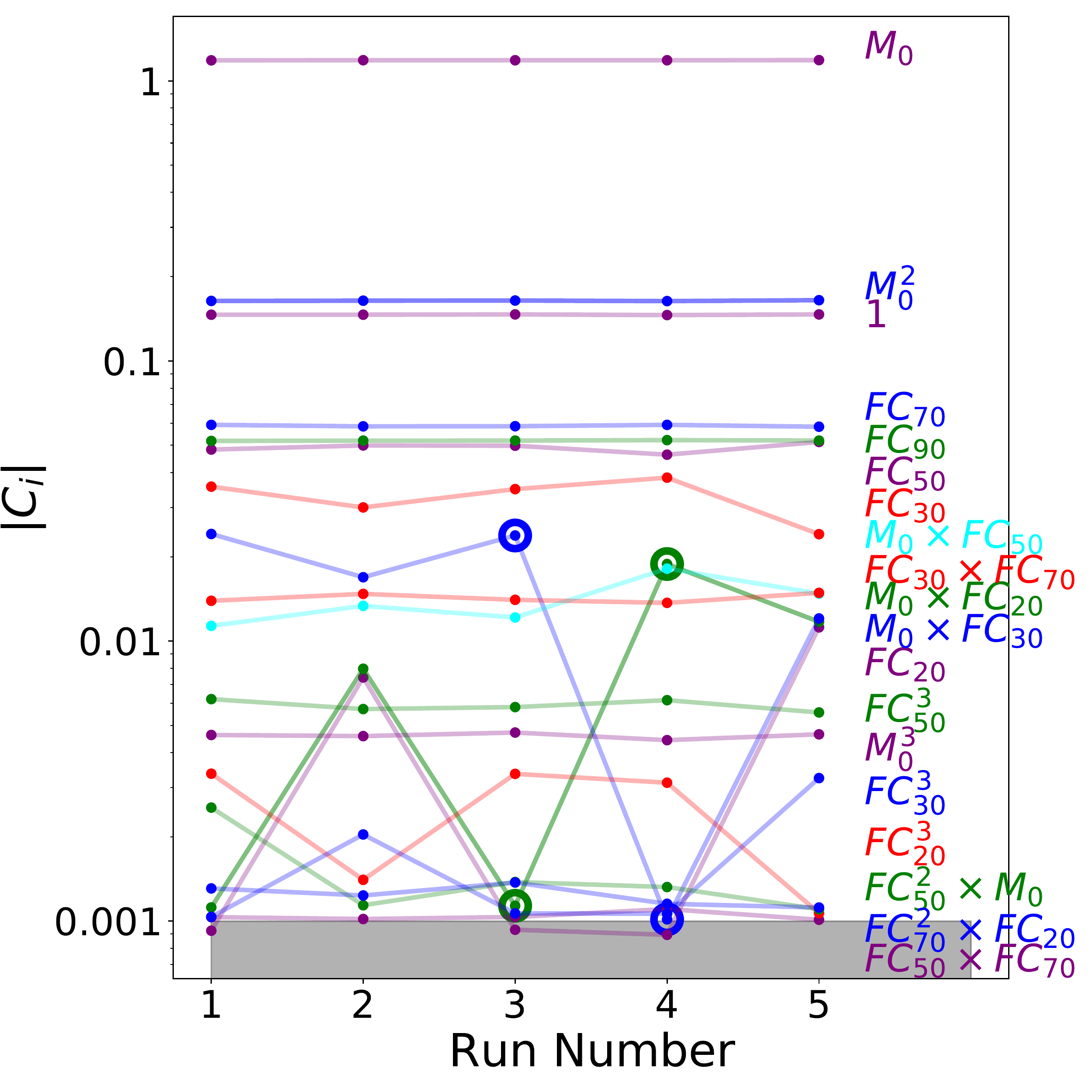}
\caption{Best fit absolute values of coefficients $C_i$ for the formation criterion model using 5 different initial positions.  
The lines connect coefficients that survived in at least one model, with the right hand key indicating which function they refer to. The colour coding of the lines is only there to help to differentiate between them. Blue and green circles correspond to the coefficients associated with $M_0 \times \mathbf{FC}_{30}$ and $M_0 \times \mathbf{FC}_{20}$ functions in runs 3 and 4 respectively. They are highlighted  as an example of correlated variables associated with different local minima. The grey area represents the cutoff value $\nu$.}
\label{DiffIP}
\end{figure}

To test how strong an effect the initial starting point is, we perform five different minimizations of the formation criterion model using 5 distinct starting points in the minimsation algorithm. We set $\lambda_\text{min}=932$, which is the optimised value found by running our methodology with our standard configuration. The initial point in the parameter space $C_i$ is varied to random values between the five runs and is the only feature that is different between runs.
Fig.~\ref{DiffIP} shows the best fit $C_i$ coefficients obtained using 5 different sets of initial positions. All models have an equivalent accuracy with a RMSE within the range $0.249\pm0.001$. Three out of the five models use 19 parameters and the remaining two use 18. All resulting models have equivalent accuracy and simplicity and we can not select one as being significantly better than the rest.

We can tell that the most significant coefficients (i.e.\ those with a larger $C_i$) are kept constant amongst all runs, similarly there is a large subset of coefficients that are not necessary in any of the models. However, there is a subset of parameters that are interchangeable between different models. An example of this is shown by the green and blue circles, which correspond to the coefficients associated with $M_0 \times \mathbf{FC}_{30}$ and $M_0 \times \mathbf{FC}_{20}$ functions in runs 3 and 4 respectively. Both runs are very similar in almost every parameter, except that run 3 gives a very important role to the $M_0 \times \mathbf{FC}_{30}$ function and almost discards the $M_0 \times \mathbf{FC}_{20}$ function, while run 4 does the opposite. This indicates that both parameters are correlated with each other and that our methodology can choose one or the other and still come up with equivalent solutions.

\begin{figure}
\includegraphics[width=\linewidth]{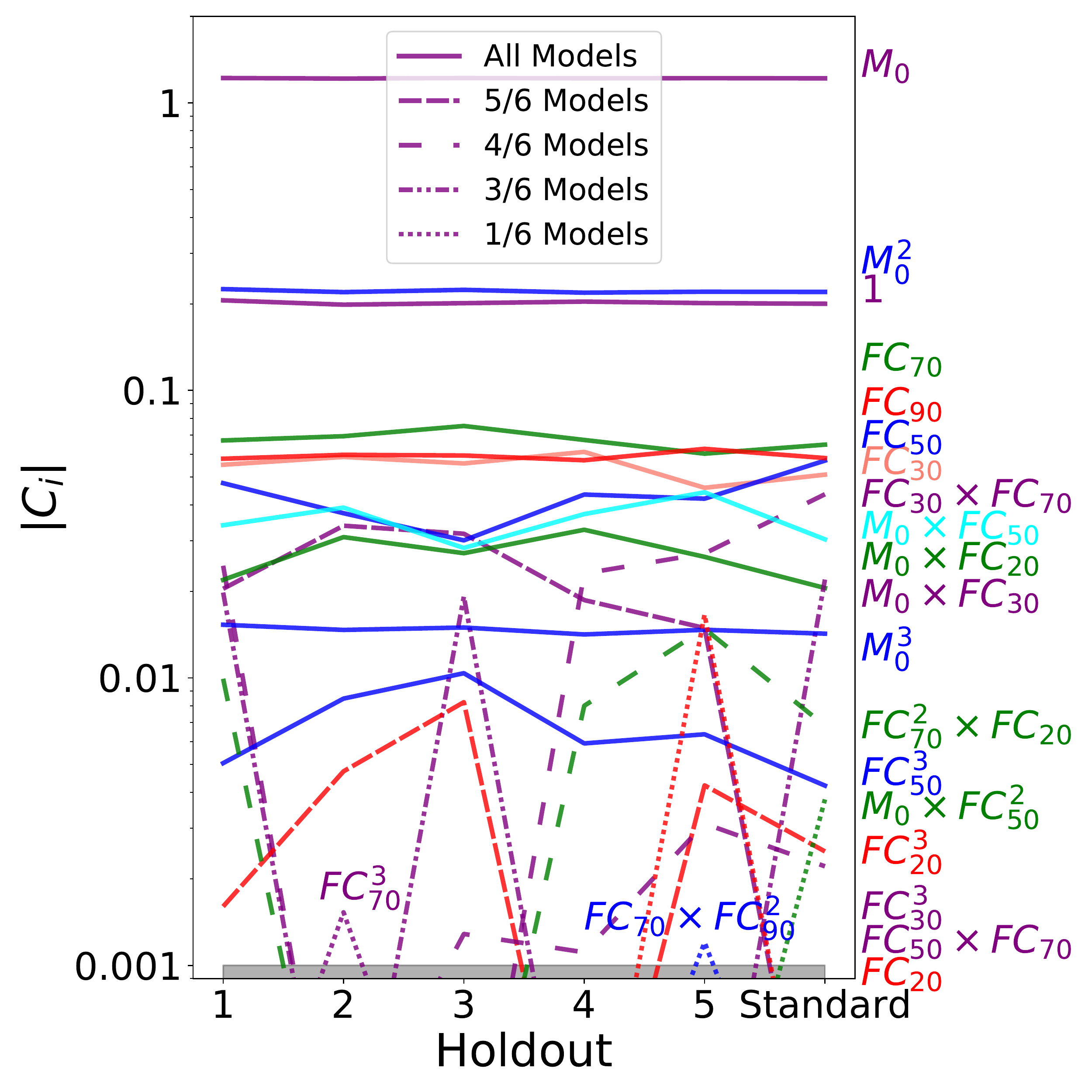}
\caption{Best fit absolute values of the coefficients $C_i$ for the formation criterion model using six different holdouts, with the right most one corresponding to the standard holdout set used throughout this paper.
The lines connect coefficients that survived in at least one model, with the right hand key indicating which function they refer to. The line style indicates how often a given coefficient was kept by the best fit model (as indicated by the key). The colour coding of the lines is only there to help to differentiate between them. Each run uses $\%15$ of the data as holdout set, each of which are disjoint from each other.
The resulting models, which have similar accuracy (RMSE=$0.166\pm0.004$), select a somewhat different subsets of surviving coefficients $C_i$, with the most important ones remaining the same and the less important ones often exchanged for comparable ones. See text for further discussion.}
\label{DiffHold}
\end{figure}

To test the variance of our methodology, we make six independent runs of the formation criterion model, varying only the holdout set, the data that is kept outside of the model fitting process. One of the holdout sets is our standard holdout, used throughout the paper. The other five correspond to five independent subsets of the training set with the same amount of points that the standard holdout set: the six independent holdouts considered have each $15\%$ of the whole data set. The RMSE of the 6 resulting models are [0.167,~0.170,~0.169,~0.162,~0.162,~0.167] and they have [16,~14,~15,~15,~18,~17] surviving coefficients each respectively. Therefore all six models have similar accuracy and comparable simplicity. 
Fig.~\ref{DiffHold} shows the variations in the resulting $C_i$ coefficients that survived in at least one of the six models. Solid line are used for the eleven coefficients that survived in all of six runs. This means that on average two thirds of all coefficients are the same irrespective of the specific holdout data set used. We note that the numerical values of those eleven coefficients are often of similar amplitude in all runs. Of the remaining coefficients, two are present in five of six models and a further two in four of six. Hence there are 15 coefficients present in nearly all six models, indicating how robust our algorithm is to changes in the holdout set used. We note that some of the other coefficients found in some runs are likely correlated with those ones and are sometimes present but discarded in at least half of the runs.

\section{Results}
\label{results}
\label{result_ratios}

We now present the results of our four models defined in  Section~\ref{Experiment_list}, i.e.: (i) Mass ratio, (ii) Formation criterion, (iii) Mass ratio and $\vec{\mathbf{j}}$ and (iv) Formation criterion and $\vec{\mathbf{j}}$. The specific surviving coefficients $C_i$ selected by each of the models are presented in Table~\ref{Results_table}, where coefficients are reported in standardized space. They can not be used directly to model the actual data, which needs to be transformed using Eq.~\ref{Normalisation}. The standardised space is defined by the mean and the standard deviation of the logarithm of the stellar mass of galaxies and of the dependent variables $\vec{z}'_i$, which are shown in~\ref{Normalisation_table}.

A striking feature of models (iii) and (iv), the two models with $\vec{\mathbf{j}}$, is that the algorithm does not select any specific angular momentum parameters in either of them. In fact the selected parameters of model (ii) and (iv) are almost identical. 
While there are some small differences between the coefficients chosen in models (i) and (iii), namely model (iii) selects two extra parameters, and the values of some of the common parameters are slightly different. These difference are consistence with the variance of the methodology reported in Section~\ref{sec:uncertainty}.
This indicates that the contribution to the accuracy of the model after including the angular momentum parameters is negligible: the sparse regression method found that no angular momentum parameters contributed additional information necessary to describe the SMHM relation that was not already provided using the rest of the parameters. This suggests that any correlation between the specific angular momentum history of a galaxy and that of its host halo should be the consequence of a correlation between the mass and specific angular momentum formation histories of host halos.

\begin{table*}

\begin{tabular}{|c||c|c||c|c||c|c|||c|c|}
 \hline

Models: &  \multicolumn{2}{c}{(i) mass ratio} & \multicolumn{2}{c}{(ii) formation criterion} & \multicolumn{2}{c}{(iii) mass ratio and $\vec{\mathbf{j}}$} & \multicolumn{2}{c}{(iv) formation criterion and $\vec{\mathbf{j}}$} \\
\hline
 N & Coefficient & Value & Coefficient & Value& Coefficient & Value& Coefficient & Value \\
\hline
1 & 1 & 0.120 & 1 & 0.200 & 1 & 0.139 & 1 & 0.200 \\ 
2 & $M_0$ & 1.22 & $M_0$ & 1.217 & $M_0$ & 1.22& $M_0$ &1.217 \\ 
3 & $M_{0.18}/M_0$ & 0.0335 & $\mathbf{FC}_{30}$ & 0.0509 & $M_{0.18}/M_0$ & 0.0334& $\mathbf{FC}_{30}$ & 0.0509 \\ 
4 & $M_{0.37}/M_0$ & 0.0469 & $\mathbf{FC}_{50}$ & 0.0567 & $M_{0.37}/M_0$ & 0.0465& $\mathbf{FC}_{50}$ & 0.0570 \\ 
5 & $M_{0.62}/M_0$ & 0.0662 & $\mathbf{FC}_{70}$ & 0.0648 & $M_{0.62}/M_0$ & 0.0671& $\mathbf{FC}_{70}$ & 0.0647 \\ 
6 & $M_{1}/M_0$ & 0.0410 & $\mathbf{FC}_{90}$ & 0.0582 & $M_{1}/M_0$ & 0.0412&  $\mathbf{FC}_{90}$ & 0.0582\\ 
7 & $M_{1.26}/M_0$ & 0.0312 & $M_0^2$ & -0.220 & $M_{1.26}/M_0$ & 0.0284& $M_0^2$ & -0.220 \\ 
8 & $M_{1.74}/M_0$ & 0.0578 & $M_0\times\mathbf{FC}_{20}$ & -0.0205 & $M_{1.74}/M_0$ & 0.0572 &  $M_0\times\mathbf{FC}_{20}$ & -0.0206 \\ 
9 & $M_{2.48}/M_0$ & 0.0545 & $M_0\times\mathbf{FC}_{30}$ & -0.0220 & $M_{2.48}/M_0$ & 0.0357& $M_0\times\mathbf{FC}_{30}$ & -0.0219 \\ 
10 & $M_0^2$ & -0.172 & $M_0\times\mathbf{FC}_{50}$ & -0.0304 & $M_{3.02}/M_0$ & 0.0162& $M_0\times\mathbf{FC}_{50}$ & -0.0304 \\
11 & $(M_{1}/M_0)^2$ & 0.00936 & $\mathbf{FC}_{30}\times\mathbf{FC}_{70}$ & -0.0435 & $M_0^2$ & -0.215& $\mathbf{FC}_{30}\times\mathbf{FC}_{70}$ & -0.0435\\ 
12 & $(M_{1.74}/M_0)^2$ & 0.0136 & $M_0^3$ & 0.0143 & $(M_{1}/M_0)^2$ & 0.00928& $M_0^3$ & 0.0143 \\ 
13 & $(M_{2.48}/M_0)^2$ & 0.00747 & $\mathbf{FC}_{20}^3$ & 0.00249 & $(M_{1.74}/M_0)^2$ & 0.0130& $\mathbf{FC}_{20}^3$ & 0.00249 \\ 
14 & $(M_{3.02}/M_0)^2$ & -0.00237 & $\mathbf{FC}_{30}^3$ & 0.00221 & $(M_{2.48}/M_0)^2$ & 0.00706& $\mathbf{FC}_{30}^3$ & 0.00221 \\ 
15 & $(M_{3.98}/M_0)^2$ & 0.00265 & $M_0 \times \mathbf{FC}_{50}^2$ & 0.00378 & $(M_{3.02}/M_0)^2$ & -0.00326& $M_0 \times \mathbf{FC}_{50}^2$ & 0.00378 \\ 
16 & $M_0\times (M_{0.62}/M_0)$ & -0.0169 & $\mathbf{FC}_{50}^3$ & 0.00422 & $(M_{3.98}/M_0)^2$ & 0.00674& $\mathbf{FC}_{50}^3$ & 0.00422 \\ 
17 & $M_0\times (M_{1.26}/M_0)$ & -0.0114 & $\mathbf{FC}_{70}^2\times\mathbf{FC}_{20}$ & 0.00682 & $M_0\times (M_{0.62}/M_0)$ & -0.0169& $\mathbf{FC}_{70}^2\times\mathbf{FC}_{20}$ & 0.00682 \\ 
18 & $M_0\times (M_{1.74}/M_0)$ & -0.0139 &  &  & $M_0\times (M_{1.26}/M_0)$ & -0.01171& & \\ 
19 & $M_0\times (M_{3.02}/M_0)$ & -0.0236 &  &  & $M_0\times (M_{1.74}/M_0)$ & -0.01244& & \\ 
20 & $M_0^2\times (M_{3.98}/M_0)$ & -0.00415 &  &  & $M_0\times (M_{3.02}/M_0)$ & -0.0348& & \\ 
21 & & &  &  & $M_0^3$ & 0.0142&  & \\ 
22 & &  &  &  & $M_0^2\times (M_{3.98}/M_0)$ & 0.00643&  & \\ 

\end{tabular}
\caption{Parameters and their values as selected by each of the four models of Section~\ref{Experiment_list}. Neither the mass ratio and $\vec{\mathbf{j}}$ model nor the formation criterion and $\vec{\mathbf{j}}$ model used any specific angular momentum parameters as part of their final coefficients. The formation criterion based models, i.e.\ models (ii) and (iv), are virtually identical with only very minor differences in some of the coefficient values. We note that the parameters in this table are all quoted in the standardised space, i.e.\ where all dependent variables have made used of Eq.~\ref{Normalisation}. Parameters are shown to three significant figures, which we find are enough to make the RMSE accurate to the fourth significant figure.}
\label{Results_table}
\end{table*}

\begin{figure*}
\includegraphics[width=50mm,height=50mm]{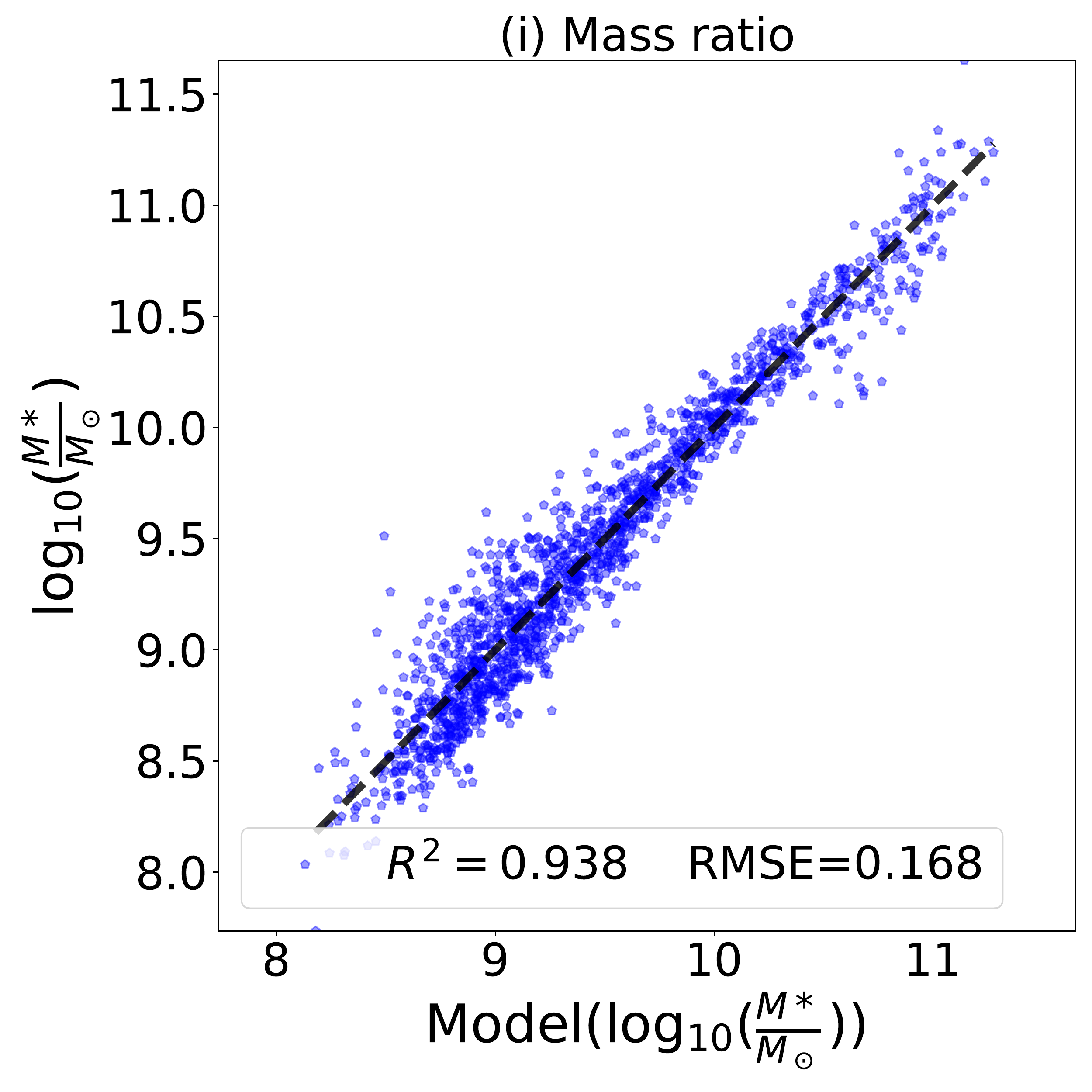}
\includegraphics[width=50mm,height=50mm]{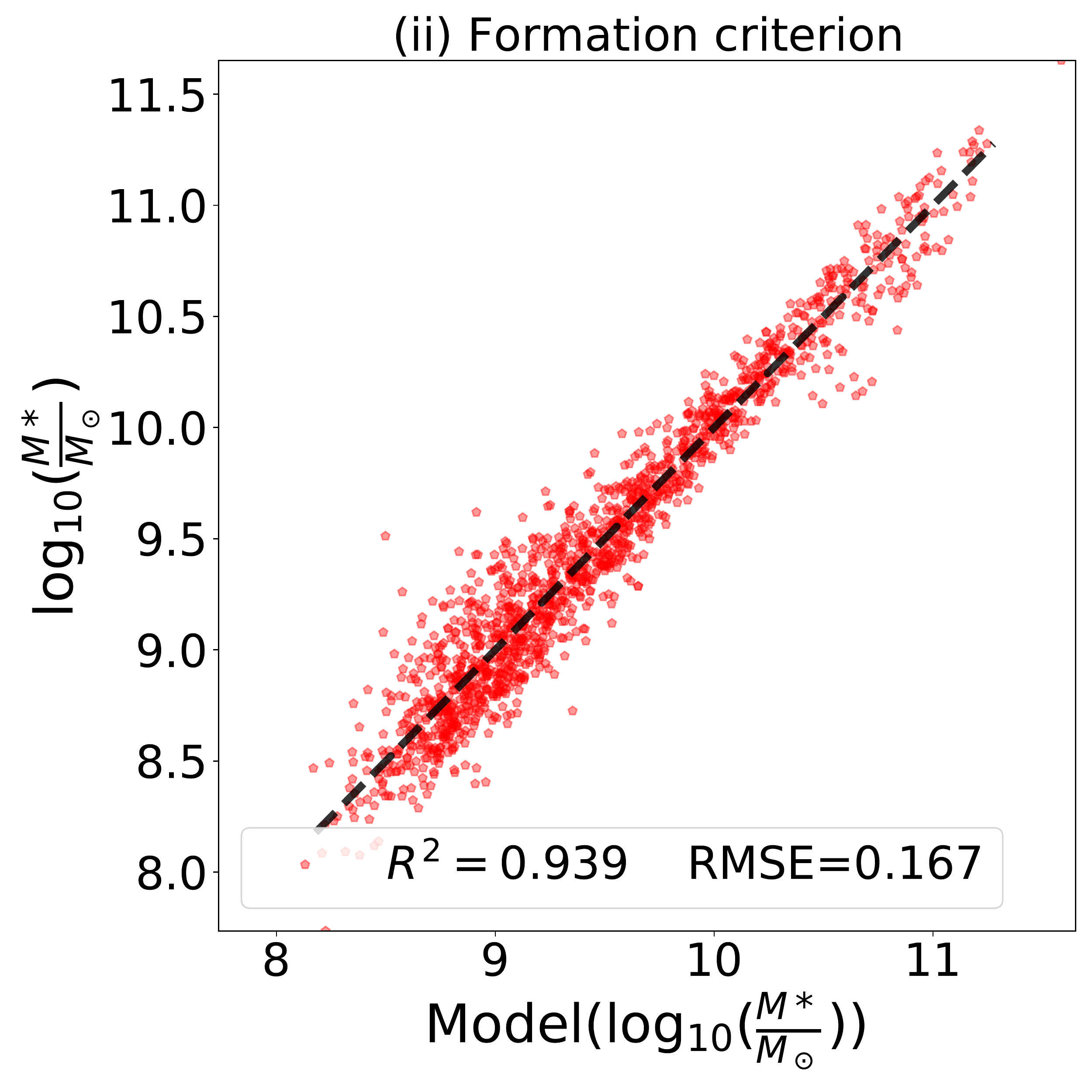}
\includegraphics[width=50mm,height=50mm]{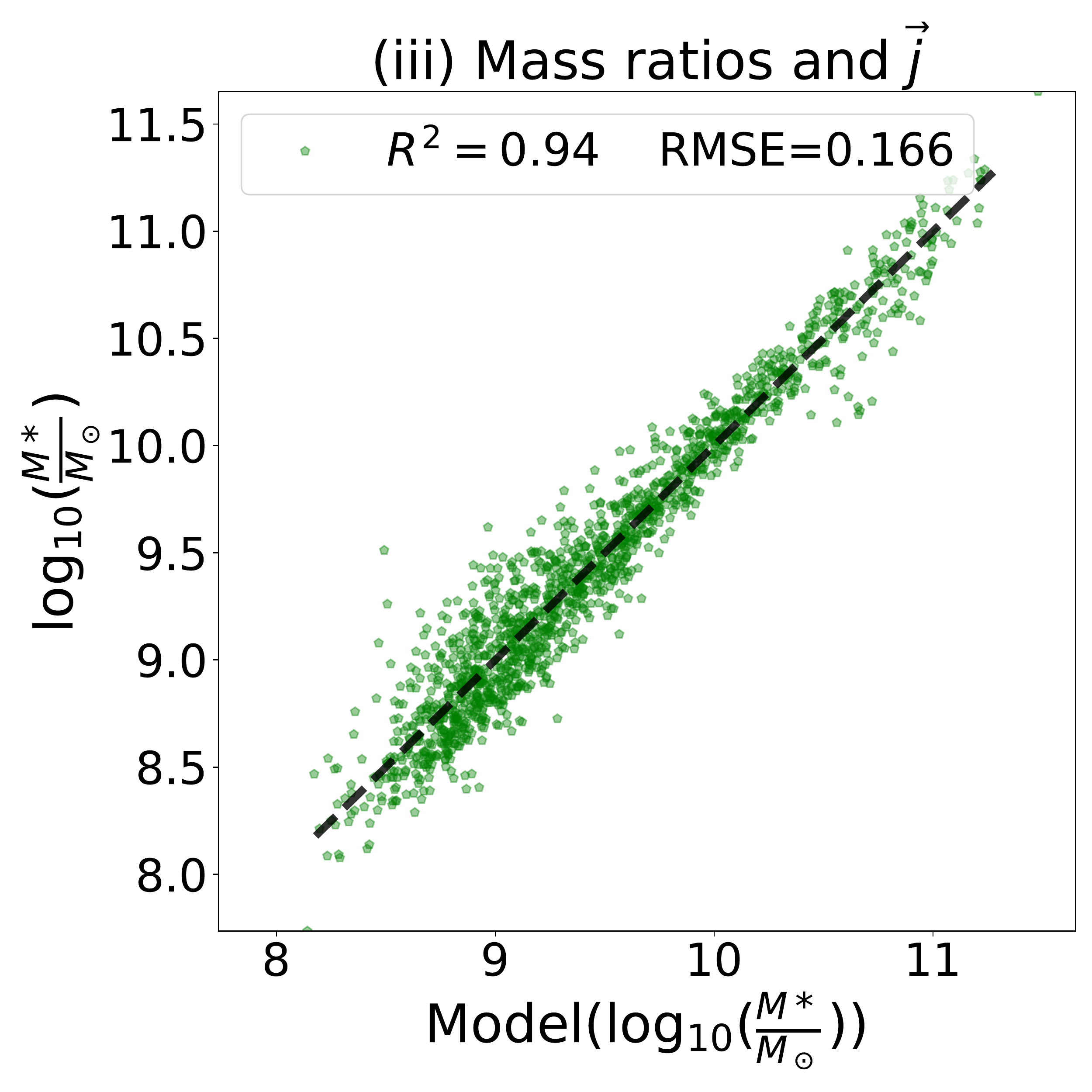}
\caption{Comparison between the stellar mass predicted by the models and its actual value in the EAGLE simulation for all galaxies in the holdout set. Left, centre and right panels correspond to the mass ratio, the formation criterion, and the mass ratio and $\vec{\mathbf{j}}$ models respectively (as indicated in the header of each panel). The closer each point is to the one-to-one relation (black dashed lines), the more accurate the model prediction is. The value of the RMSE and the $R^2$ statistic are included for each model. In general the three models have equivalent accuracy. As the formation criterion and $\vec{\mathbf{j}}$ model is virtually identical to the formation criterion model (see Table~\ref{Results_table} for parameter values), we included only the latter one.}
\label{ModelvsData}
\end{figure*}

Fig.~\ref{ModelvsData} shows the predicted values of the stellar mass for all galaxies in the holdout set for three models (omitting model (iv) as it is so similar to model (ii)) compared to their real values in the EAGLE simulation. The closer a point is to the one-to-one line (black dashed line), the better the model predicted its value. We also include the RMSE of each model, as given by Eq.~\ref{RMSE}. 

A different estimate of the goodness of a fit is the $R^2$ statistic, which determines the amount of the variation in $\vec{y}$ that can be explained by a model\footnote{$R^2$ estimators should be considered with caution as they are easily biased by inaccurate estimations of $\sigma_{y_\alpha}$ and can have deceivingly small (or large) values. They should be used as reference only. We also include RMSE errors as goodness of fit estimators, which are far more robust.}:

\begin{equation}
\label{R2}
\mathbf{R}^2=1-\frac{\mathbf{RMSE}^2}{\sigma_{y}^2},  
\end{equation}
where $\sigma_{y}$ is the standard deviation of $\vec{y}$. The usefulness of the $R^2$ comes from being intuitive to interpret: the closer to one the $R^2$ of a model is, the more accurate it is. 

Both the RMSE and $R^2$ statistics show that the three models have very similar accuracy. The formation criterion model is slightly simpler than both mass ratio models, as the former has 17 free parameters, compared with 20 and 22 from the two mass ratio models. This suggests that the formation criteria parameters, $\mathbf{FC}_i$, are slightly more efficient at summarising the halo mass information than the mass ratio parameters, ($M_z/M_0$). 

\begin{table*}
\begin{tabular}{|c||c|c|c|c|c|c|c|}
 \hline

 &  $\log_{10} M^*/M_{\odot}$ & $M_0'$ & $\mathbf{FC}_{20}'$ & $\mathbf{FC}_{30}'$ & $\mathbf{FC}_{50}'$ & $\mathbf{FC}_{70}'$ & $\mathbf{FC}_{90}'$ \\
\hline
$\mu$ & 9.460 & 11.59 &  2.748 &  2.181 &  1.419 &  0.8794 &  0.3949 \\
$\sigma$ & 0.6764 & 0.4723 & 0.7834 & 0.7389 & 0.5731 & 0.4125 & 0.2581 \\ 

\end{tabular}
\begin{tabular}{|c||c|c|c|c|c|c|c|c|c|}
\hline
& $(M_{0.18}/M_0)'$ & $(M_{0.37}/M_0)'$ & $(M_{0.62}/M_0)'$ & $(M_{1.0}/M_0)'$ & $(M_{1.26}/M_0)'$ & $(M_{1.74}/M_0)'$ & $(M_{2.48}/M_0)'$& $(M_{3.02}/M_0)'$& $(M_{3.98}/M_0)'$ \\
\hline
$\mu$ & -0.03263 & -0.06793 &  -0.1233 &  -0.2254 &  -0.2973 &  -0.4352 &  -0.6548&  -0.8086 &  -1.070 \\
$\sigma$ & 0.06933 & 0.09271 & 0.1147 & 0.1461 & 0.1660 & 0.2008 & 0.2468 &  0.2760 &  0.3254 \\ 

\end{tabular}
\caption{Normalization parameters used for the stellar mass and the DM halo variables defined in section~\ref{Input_param} and considered by our models. 
The $\mu$ and $\sigma$ rows correspond to the mean and standard deviation of the variables respectively and are used in Eq.~\ref{Normalisation} to standardize the units of the variables considered.}
\label{Normalisation_table}
\end{table*}

\subsection{Comparison with simpler models}
\label{Comparison}
While the LASSO approach uses only a fraction of the full set of available regression terms, the models it selects are still relatively complex and include non-linear combinations of terms characterising the formation history. In this section, we compare our results to simpler models. Specifically, we compare the formation criterion model from the last section with the following two models:

\begin{itemize}
    \item The first model is a third-order polynomial fit of the SMHM relation. This model includes the terms $1$, $M_0$, $M_0^2$, and $M_0^3$. We label this model as $M_0^3$, with all four coefficients selected by our LASSO method\footnote{The coefficients are $C(1)=0.179$, $C(M_0)=1.16$, $C(M_0^2)=-0.205$, $C(M_0^3)=0.0152$, when quoted in the standardised space.}.
    
    \item  Our second model is similar to the one presented in Equation (9) of \cite{Matthee_2016}. We include all terms of $M_0$ up to the third order and all linear terms of $FC_{50}$. More specifically, the eight possible terms are $1$, $M_0$, $M_0^2$, $M_0^3$, ${\bf FC}_{50}$, $M_0 \times {\bf FC}_{50}$, $M_0^2 \times {\bf FC}_{50}$, and $M_0^3 \times {\bf FC}_{50}$. We did not use the model presented in \cite{Matthee_2016} directly because of small differences in the calibration redshift and in the methodology used for selecting and processing the EAGLE data sets. We label this model as ($M_0^3$ $\&$ $FC_{50}$), with six coefficients selected by our LASSO method\footnote{The coefficient are $C(1)=0.156$, $C(M_0)=1.22$, $C({\bf FC}_{50})=0.199$, $C(M_0^2)=-0.169$, $C(M_0 \times {\bf FC}_{50})=-0.274$, $C(M_0^3 \times {\bf FC}_{50})=-0.00402$, in standardised space. The remaining terms were discarded by our LASSO methodology.}.  We have tested the prediction of this model against the predictions of \cite{Matthee_2016} \footnote{Following a discussion with the authors, we identified an issue with the model in the way it was reported in the paper. The corrected model description is $\log_{10}(M^*)=\alpha-e^{\beta \, M_0^D+\gamma}-(a \, FC_{50}+b)$ where $M_0^D = M_0 - 12$,  $a = 0.15048 + 0.21517 \, M_0^D + 0.06412 \, (M_0^D)^2 - 0.07217 \, (M_0^D)^3$, $b = 0.20632 - 0.43077 \, M_0^D + 0.25277 \, (M_0^D)^2 + 0.34500 \, (M_0^D)^3$ and $\alpha$, $\beta$ and $\gamma$ are constants which values are given in Table~2 of \cite{Matthee_2016}} and find that the models are comparable.
\end{itemize}

\begin{figure}
\includegraphics[width=85mm,height=85mm]{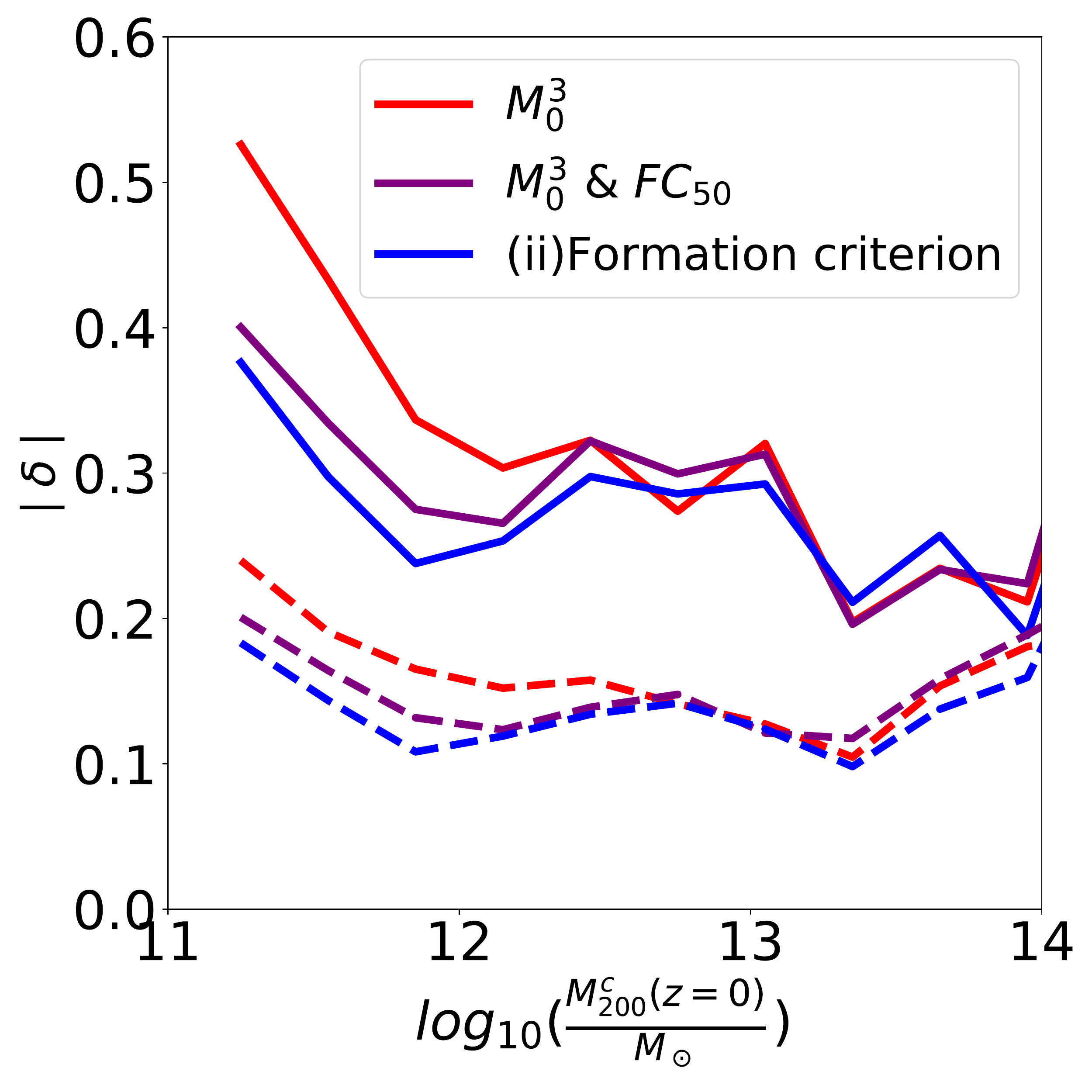}
\caption{Comparison of the accuracy of the models discussed in Section~\ref{Comparison}, as traced by the $\delta$ error (also defined in \S\ref{Comparison}) as a function of the present day halo mass. Dashed and solid lines correspond to the 68$^{\rm th}$ and 95$^{\rm th}$ percentiles of the absolute value of error distribution.
}
\label{percentile_errors}
\end{figure}

As the models grow in complexity, their prediction of the stellar mass becomes more accurate, a way of quantifying this is by looking at the RMSE of our data set. 

The $M_0^3$ model has a RMSE of 0.225 when estimated with stellar mass units\footnote{In this sub-section, the RMSE is expressed in natural units, i.e.\ the logarithm of the stellar mass. This results in RMSE values which are more natural to understand, as here an RMSE of 0.2 implies that the mean error is $0.2 \log_{10}(M^*/M_{\odot})$. We note that the RMSE depend on the parametrisation, which throughout the rest of this work is the one defined using standardised units (see Eq.~\ref{RMSE}).}.  We obtain very similar results looking both at the holdout set and the whole dataset. For the $M_0^3$ \& ${\bf FC}_{50}$ model, the stellar mass RMSE drops to 0.181, while it is 0.166 for the formation criterion model. Assuming that contributions from the different terms can be added in quadrature, this shows that 32\% of the variance of the $M_0^3$ model is explained by including linear terms in ${\bf FC}_{50}$, while the more complex model selected by the LASSO process explains a further 10\% of the variance, a modest but significant improvement.

This suggests that the biggest improvement on the SMHM residuals (modeled by $M0^3$) comes from the linear terms of ${\bf FC}_{50}$, the higher-order terms of ${\bf FC}_{50}$ and the terms corresponding to other formation criteria make a smaller but significant correction to the predicted stellar mass.

To explore the improvement of the model further, we define for each galaxy the error of a model as the difference between the actual stellar mass and the predicted one, or more precisely:
\begin{equation}
\delta=\log_{10}\left( \frac{M^*}{M^*_p(\vec{C},{\bf X'})}\right)
\end{equation}
where $M^*_p$ corresponds to the model predicted stellar mass of a galaxy of stellar mass $M^*$.
Fig.~\ref{percentile_errors} shows the 68$^{\rm th}$ and 95$^{\rm th}$ percentile ranges of $\mid \delta \mid$ as a function of halo mass for the reference formation criterion model (blue lines), and the $M_0^3$ $\&$ ${\bf FC}_{50}$ and $M_0^3$ models (purple and red lines respectively). The plot shows that the differences between the three models are most significant at small halo masses, while at halo masses larger than $\sim 10^{12.5} M_{\odot}$, all models are comparable. This suggests that galaxies in smaller halos are more readily explained by evolutionary effects (correlated with ${\bf FC}_p$ parameters), while the scatter in larger galaxies is perhaps more strongly influenced by stochastic baryonic processes, such as black hole accretion, that cannot be modelled using the halo mass history alone. This is in agreement with \cite{Matthee_2016} that found no correlation between the scatter of the SMHM relation and formation time for halo masses larger than $\sim 10^{12.5} M_{\odot}$.

Rather than restricting, by hand, the choice of functions to terms that are linear in ${\bf FC}_{50}$, we can of course ask the LASSO methodology to simplify the formation criterion model, trading off an increase in variance for a reduction in complexity. It should be remembered, however, that this model will not provide optimal predictions for the stellar mass in a RMSE sense. We shift the balance to reduce complexity by increasing the penalty parameter $\lambda$ of Eq.~\ref{equationtomin}.  As can be seen in Fig.~\ref{lamvsRMSE}, using a penalty $\lambda$ three times larger than the one selected by the LASSO algorithm generates a model that is comparable to model $M_0^3$ $\&$ ${\bf FC}_{50}$ in terms of the RMSE and number of surviving terms. The terms retained by the model are: $1$, $M_0$, ${\bf FC}_{50}$, ${\bf FC}_{70}$, ${\bf FC}_{90}$, ${\bf FC}_{20}^3$, ${\bf FC}_{30}^3$, ${\bf FC}_{50}^3$, with coefficients 0.0538, 1.13, 0.0315, 0.0534, 0.0242, 0.00590, 0.0104, 0.0108 respectively. Interesting this model prefers to characterise the formation histories of the haloes more precisely rather than to mix terms depending on halo mass and formation time.

\subsection{Interpretation}
\label{Interp}

The goal of this work is to make a model that is accurate and also explainable. With this in mind, we now try to give a physical interpretation to some of the terms kept in our model.

By looking at Table~\ref{Results_table}, we conclude that in general surviving parameters in all models can be divided into four different groups:
\begin{itemize}
    \item[1.] Terms forming a third order polynomial of $M_0$. Namely the terms $1$, $M_0$, $M_0^2$, $M_0^3$.
    \item[2.] Terms forming third order polynomials of the other dependent variables that are correlated with the mass at $z>0$. Namely, terms of the form $M_z/M_0$, $(M_z/M_0)^2$ and $(M_z/M_0)^3$ for the mass ratio models with and without $\vec{\mathbf{j}}$, and terms of the form $\mathbf{FC}_p$, $\mathbf{FC}_p^2$ and $\mathbf{FC}_p^3$ for the formation criterion models with and without $\vec{\mathbf{j}}$.
    \item[3.] Terms corresponding to the product of $M_0$ and either $M_z/M_0$ for the mass ratio models (i) and (iii) or $\mathbf{FC}_p$ for the formation criterion models (ii) and (iv).
    \item[4.] Other terms corresponding to higher order combinations of crossed terms, which are more challenging to provide a physical interpretation of.
\end{itemize}

The terms in group (1) correspond to a direct modelling of the SMHM relation. Let us call $P^3(z=0)$ the polynomial built with the terms in group (1) and their associated coefficients, $C_1^i$:  
\begin{equation}
\label{z0_poly}
P^3(z=0)=C_1^0+C_1^1M_{0}+C_1^2M_{0}^2+C_1^3M_{0}^3
\end{equation}
In order to compare our model stellar mass predictions with the EAGLE stellar masses, we transform our model from the standardised units to stellar mass units: 
\begin{equation}
\label{z0_poly_norm}
P'^3(z=0) = P^3(z=0) \, \sigma(\log_{10}(M^*))+\mu(\log_{10}(M^*))
\end{equation}
where the stellar mass, $M^*$, is expressed in $M_{\odot}$, $\mu$ and $\sigma$ are the mean and standard deviation operators considered in Eq.~\ref{Normalisation} already.

\begin{figure*}
\includegraphics[width=85mm,height=85mm]{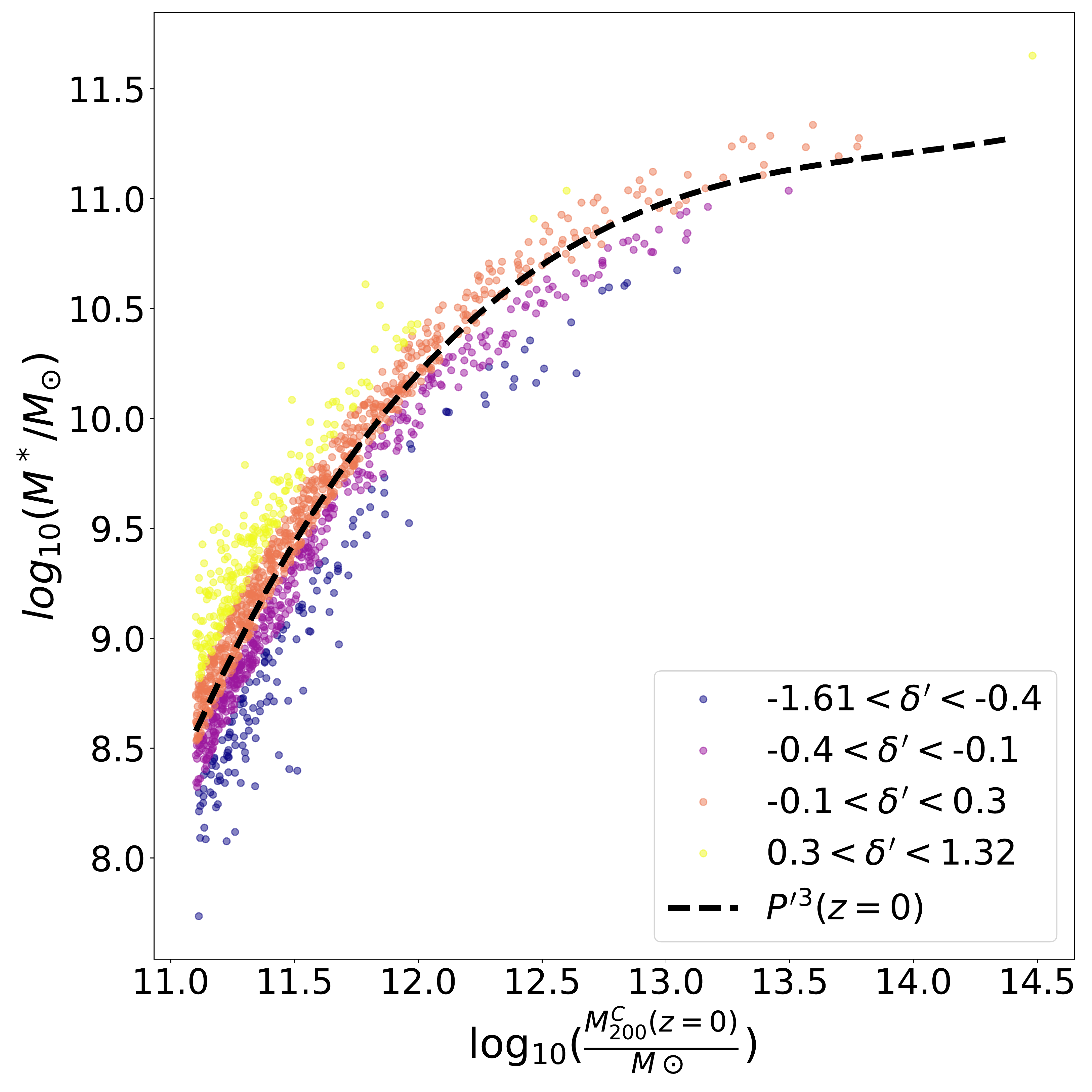}
\includegraphics[width=85mm,height=85mm]{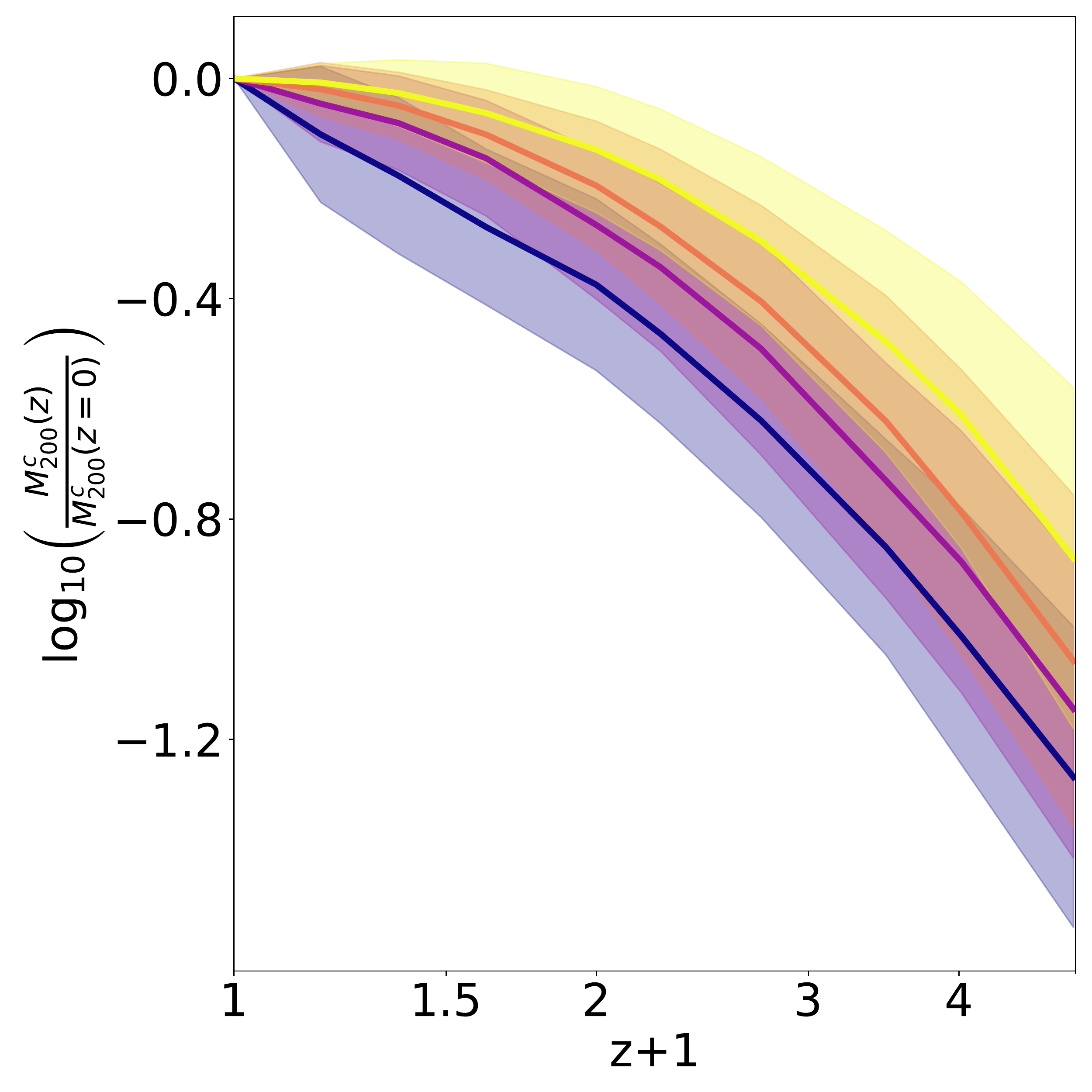}
\caption{{\bf Left panel:} The stellar mass -- halo mass relation for galaxies in our holdout set. The black dashed line shows the polynomial $P'^3(z=0)$ from Eq.~\ref{z0_poly_norm} for the formation criterion model. This line describes the trend well but there is some scatter around it. The colour coding split the galaxy sample by their residual $\delta'$ from Eq.~\ref{Residuals} into four bins, with the $\delta'$ range indicated by the key.
{\bf Right panel:} Evolution of the halo mass for each residual $\delta'$ bin, as defined in the left panel, as function of redshift. The solid lines represent the mean of the logarithm of the halo mass ratios for all galaxies in each $\delta'$ bin, with the same colour scheme as in the left panel. The shaded contours indicate the corresponding standard deviation on the mean. Galaxies with the more negative $\delta'$ residuals reside in halos that recently assembled their final halo mass, while galaxies with the more positive $\delta'$ residuals reside in halos that primarily assembled their halo mass at an earlier stage of their evolution. }
\label{Residuals_plot}
\end{figure*}

$P'^3(z=0)$ computed for the formation criterion model is shown as the black dashed curve of the left panel of Fig.~\ref{Residuals_plot}. The figure shows that $P'^3(z=0)$ provides already a good model of the SMHM relation; however, there is some scatter around it that the model does not account for. We define the residual between each galaxy and the model prediction given by $P'^3(z=0)$ as $\delta'$:
\begin{align}
\label{Residuals}
\delta' &= \log_{10}(M^*/M_{\odot})-P'^3(z=0).
\end{align}
Galaxies in Fig.~\ref{Residuals_plot} are divided into four $\delta'$ bins. The yellow bin, which is the bin with the largest $\delta'$ values, correspond to galaxies for which their stellar masses are the most under-predicted by $P'^3(z=0)$, while the blue bin contains those with the most over predicted stellar masses. The right panel of in Fig.~\ref{Residuals_plot} shows the average mass of halos in each of the four bins as a function of redshift. On average galaxies in the yellow bin live inside host halos that attained their final mass early in their evolution when the characteristic density was higher. The deeper potential well of these halos allows the creation of massive galaxies. In contrast, the galaxies in the most over-predicted $\delta'$ bin (blue) live inside host halos that only achieved their final mass very recently and therefore had a lower characteristic density for a considerable period of time, compared to halos of the same mass in larger $\delta'$ bins. This implies that there is a correlation between $\delta'$ and the mass formation history and explains why coefficients in group (2) were selected by our model. 

This conclusion is in agreement with \cite{2014MNRAS.443.3044Z}, where formation time is used to model assembly bias, and with \cite{Matthee_2016}  where formation time is found to be the most correlated parameter with $\delta'$.  We emphasize that we arrive at this conclusion by using a completely different approach, that does not require any prior knowledge of the underlying physics correlating stellar mass with halo mass and formation time.

A novel result from our model is that all terms of $\mathbf{FC}_{p}$ with $p=[20,30,50,70,90]$ are needed in the final fit. This suggests that formation time alone is not enough for our model to remove the correlation with $\delta'$, but actually tracking the different formation times at which different percentages of the final halo mass were assembled leads to more accurate models.

Our model suggests that the assembly history dependence is itself a function of the final halo mass. In order to explore this, we write the polynomial fits to each of the $\mathbf{FC}_{p}$ terms from (2), and their associated coefficients, $C_p^i$:
\begin{equation}
\label{zp_poly}
P^3(p)=C_p^0+C_p^1\mathbf{FC}_{p}+C_p^2\mathbf{FC}_{p}^2+C_p^3\mathbf{FC}_{p}^3
\end{equation}
where $p=[20,30,50,70,90]$. We define the residual $\delta_p$ as the leftover residual once we have removed contributions from all terms from groups (1) and (2), i.e.:
\begin{align}
\label{Residuals_l}
\delta_p &= y- P^3(z=0) - \sum_p P^3(p)
\end{align}
where $p=[20,30,50,70,90]$. We note that $\delta_p$ is defined in standardised space, with positive $\delta_p$ corresponding to a model underprediction and negative $\delta_p$ a model overprediction.

\begin{figure}
\includegraphics[width=85mm,height=85mm]{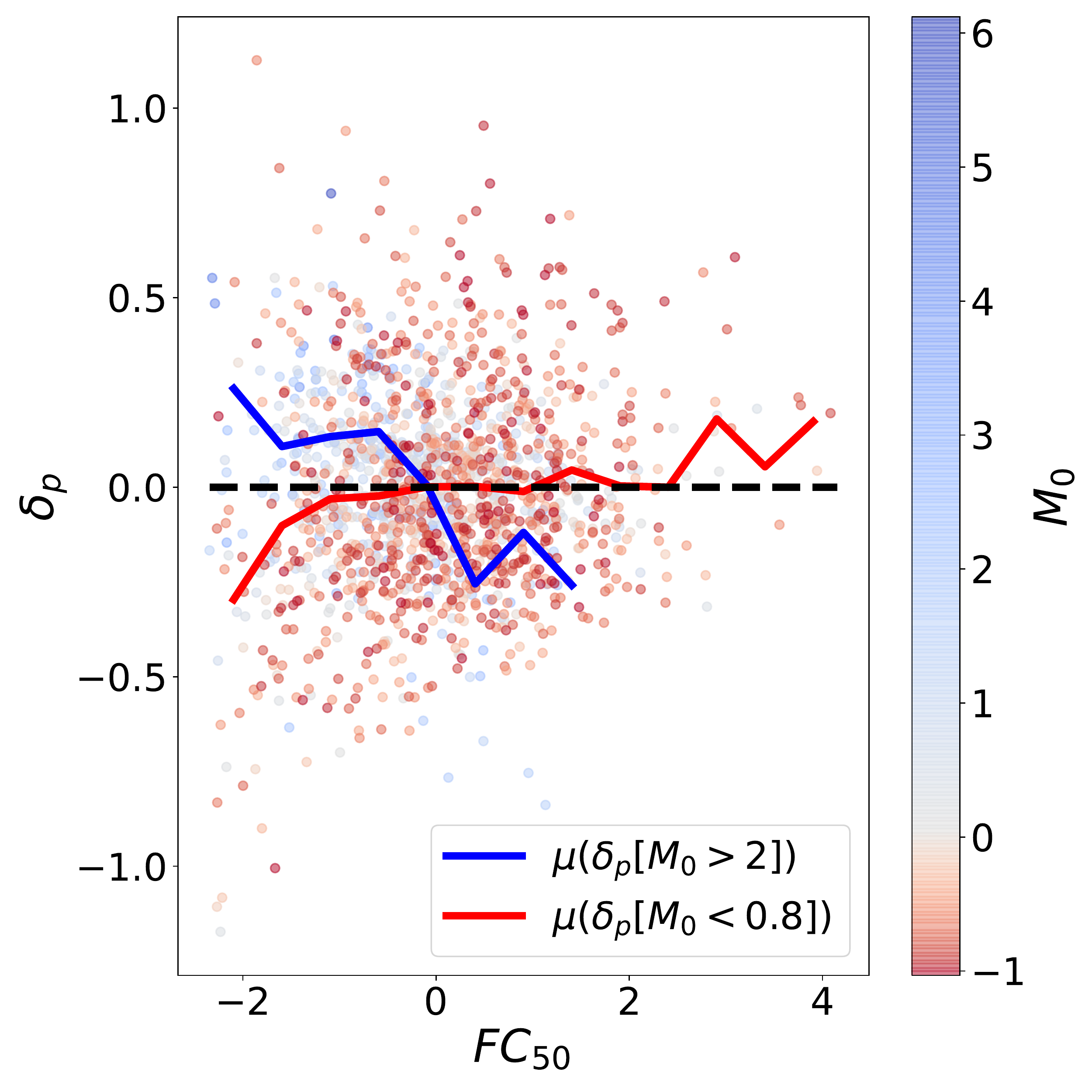}
\caption{Relation between $\mathbf{FC}_{50}$ and the residual $\delta_p$ of Eq.~\ref{Residuals_l} for all galaxies in our holdout set. $\mathbf{FC}_{50}$, which is in standardised units, maps to the redshift at which a halo acquired half of its mass. The galaxies are colour coded by $M_0$ (Eq.~\ref{M0}). The blue solid line shows the mean residual, $\mu(\delta_p)$, for galaxies in very massive halos, i.e.\ halos where $M_0>2$. Those halos are more that 2 standard deviations more massive than the mean. The red solid lines shows the mean residual, $\mu(\delta_p)$, for galaxies living in halos with very low mass ($M_0<0.8$). The blue and red lines have slopes of opposite sign, which is reflected in the presence of terms from group (3) in the solution (see \S\ref{Interp}. The plot shows that the strength of assembly bias is correlated with the final halo mass.}
\label{fc50vsdl}
\end{figure}

Fig.~\ref{fc50vsdl} shows where galaxies are in the $\mathbf{FC}_{50}$ vs $\delta_p$ plane, where $\mathbf{FC}_{50}$ (in standardized units) corresponds to the redshift when $50\%$ of the mass of a halo has been formed. 

The blue and red solid lines show the average $\delta_p$ for very massive and very small halos respectively. 
When $\mathbf{FC}_{50}$ is negative (i.e.\ smaller redshifts than the average, i.e.\ at later times), galaxies living in massive host halos tend to be overpredicted by the model (as shown by the blue line being above zero) and galaxies living in small halos tend to be underpredicted (as shown by the red line being below zero).
This shows why terms of the form $\mathbf{FC}_{p} \times M_0$, corresponding to coefficients in group (3) improve our model. The fact that the model selected terms of the form $\mathbf{FC}_{p} \times M_0$ suggests that it is not enough to model a linear relationship between stellar mass and formation time (or in our case formation criteria), but that this relation needs to be corrected by a factor that is dependent on the final halo mass. Assembly bias suggests that the stellar mass of galaxies depends on formation history, our model also suggests that this dependency is in turn dependent on the final halo mass.

\subsection{Stellar mass distribution and galaxy clustering of centrals}
\label{Discussion}

We have shown the models capability to reproduce the stellar mass of individual galaxies from the EAGLE simulation. We now discuss  our models  accuracy at reproducing other statistics from EAGLE such as the distribution of galaxy masses through the stellar mass function (SMF), and the clustering of the galaxies via 2-point correlation functions.

We consider the six realisations of the formation criterion model presented in Fig.~\ref{DiffHold} as a way of providing some uncertainty on the best fit model predictions.
Throughout this section any model comparison with EAGLE relates to comparisons with central galaxies in EAGLE, as our model only make predictions for such galaxies. 

\begin{figure}
\includegraphics[width=\linewidth]{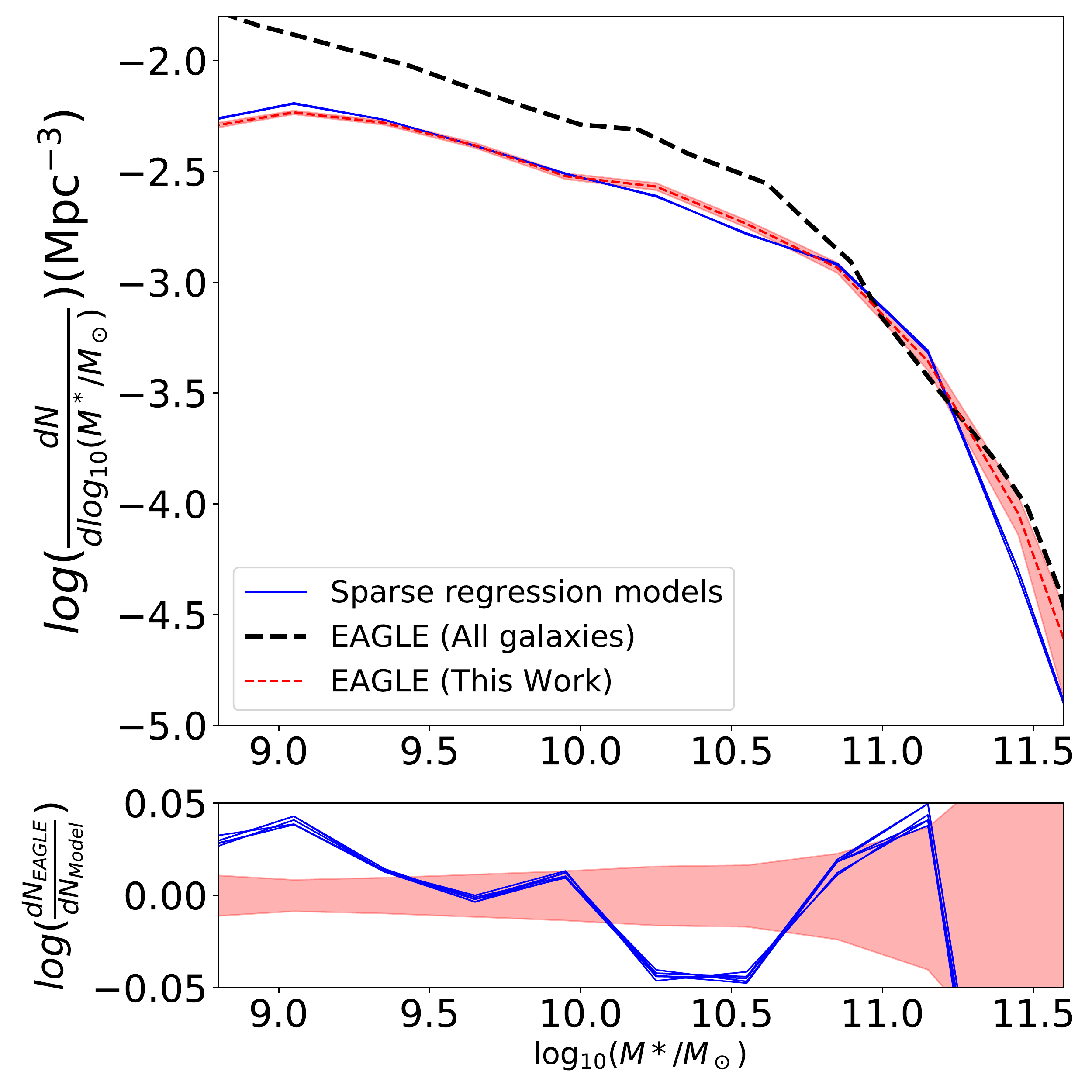}
\caption{{\bf Top panel:} 
the SMF of EAGLE galaxies used in this analysis (i.e.\ centrals) is shown in red. The SMF predicted by the six models built with our methodology is shown in blue. The red shaded area show the bootstrap errors on the SMF. For comparison, the SMF of all EAGLE galaxies is shown in black: this sample includes both centrals and satellites and does not include any halo mass cut. {\bf Bottom panel:} the ratio of predicted to actual SMFs, indicating that our models result in SMF estimates which are within 12\% of the input data on the stellar mass scales where the input data have good statistics.}
\label{Mass_Functions}
\end{figure}

\cite{2015MNRAS.450.4486F} shows that the SMF of the EAGLE hydrodynamical simulation at redshift zero agrees reasonably well with the one observed from SDSS \citep{2009MNRAS.398.2177L} and GAMA \citep{2012MNRAS.421..621B}. The red dashed line of Fig.~\ref{Mass_Functions} shows the central galaxy stellar mass function obtained from the stellar masses in our EAGLE data set.

The red shaded region is an estimate of the error due to Poisson noise within the EAGLE sample and is computed with the bootstrap method \citep{efron1979}.

The blue lines in Fig.~\ref{Mass_Functions} are the SMFs computed using the stellar masses predicted by each of our models. The predictions are so similar that it is difficult to differentiate between them, especially in the top panel. The bottom panel of Fig.~\ref{Mass_Functions} shows that the model SMFs are within 12\% of the input EAGLE SMF over most of the mass range.  At stellar masses above $\log_{10}(M^*/M_{\odot})=11.0$  the agreement of the models SMF worsens. This is likely due to the relatively small number of galaxies at this mass range in our sample (90 out of 9521). One of the many issues of including a comparatively small sample of galaxies is that the methodology has little incentive to fit them accurately as their contribution to the goodness of fit estimations is small. One possible way to improve this is to weigh their contribution more heavily than the one from galaxies in a more numerous mass range, this possibility will be explored in future iterations of this work.  

The scatter in the mass function between different models is smaller than the bootstrap error (shown as the shaded area), which suggests that the difference between the SMF of EAGLE and that of our model is not due to random sampling effects. There are notable deviations at $\log_{10}(M^*/M_{\odot})=9.0$ and $\log_{10}(M^*/M_{\odot})=10.5$. The disagreement at $\log_{10}(M^*/M_{\odot})=9.0$ is likely to be caused by selection effects, as we include a cut in halo mass which can have an effect in our model predictions at those lower stellar masses. At $\log_{10}(M^*/M_{\odot})=10.5$ the remaining residuals of the model are systematically larger and asymmetric, with the offset possibly correlated with other terms not included in our methodology. These parameters could be either other halo mass properties that we have not characterised, higher-order correlations of our input parameters, or the stochastic nature of baryonic processes. For example, feedback from super massive black holes has a highly non-linear effect on the stellar mass, either by affecting it directly, or through its influence on the baryon density inside halos \citep{bower2017, 10.1111/j.1365-2966.2012.20879.x}. Whatever the cause, characterising these asymmetric residuals remains a challenging but important problem.

As a result of the asymmetric scatter, we find that the SMFs predicted by the simpler models, $M_0^3$ and $M_0^3$ \& ${\bf FC}_{50}$ from section~\ref{Comparison}, have only minor deviations from those predicted by the full model (the formation criterion model shown in Fig.~\ref{Mass_Functions}). Although the more complex models predict more accurately the median stellar mass, all the models assume that the residuals are symmetric around this value: i.e., while the errors of a more complex model are smaller, they are not more symmetrical around their mean value. An improved treatment will have to characterise the spread of points as well as predicting a median of the relation.

The EAGLE hydrodynamical simulation has been shown to accurately reproduce the observed two point correlation function of galaxies from $1 h^{-1} \text{Mpc}$ and up to $6 h^{-1} \text{Mpc}$ \citep{10.1093/mnras/stx1263}. 

In order to test how well our model reproduces the correlation function of EAGLE galaxies, we divide the galaxies in each of our models into four stellar mass bins. We then compute the two point correlation function of galaxies in each mass bin. This is done by assigning to each model galaxy the same co-moving coordinates as that of the centre of its host halo.

\begin{figure}
\includegraphics[width=85mm,height=85mm]{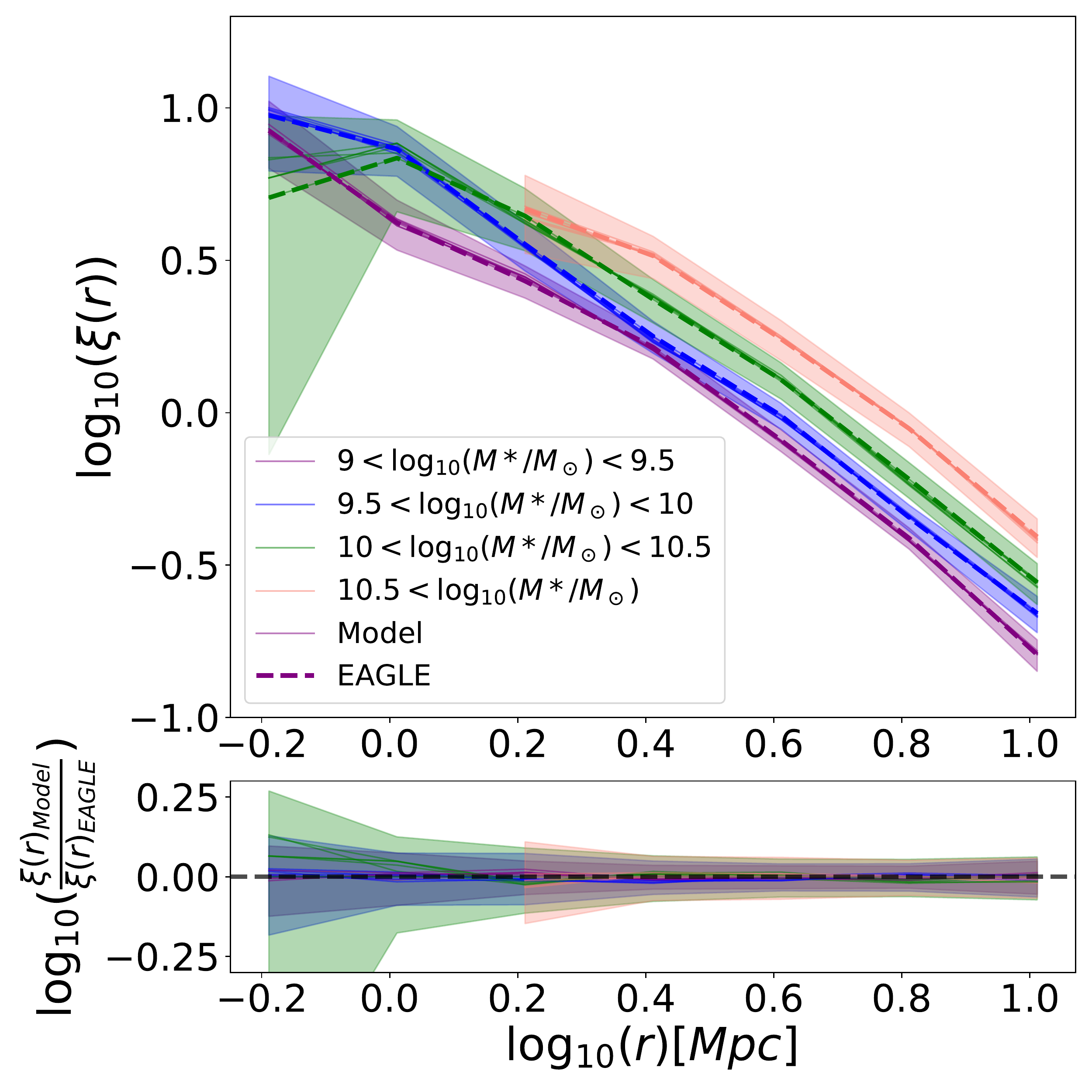}
\caption{{\bf Top panel:} correlation functions of EAGLE galaxies split into four stellar mass bins (coloured dashed lines as per key) compared to the clustering computed with our 6 models (i.e.\ 6 thin solid lines for each stellar mass bin). 
Bootstrap errors are shown on the EAGLE correlation functions.
{\bf Bottom panel:} the ratio of the predicted to the actual galaxy clustering for each stellar mass bin (same colour coding as in the upper panel). This indicates that our models result in galaxy clustering estimates split by stellar mass that agree well within the bootstrap errors with the actual clustering of EAGLE galaxies.
}
\label{CorrelationBins}
\end{figure}

Fig.~\ref{CorrelationBins} shows how the correlation functions of our models split by predicted model stellar mass compares with those obtained from the EAGLE simulation, split by the actual galaxy stellar mass. Each colour corresponds to a different mass bin, with each of our six models and for each stellar mass bin shown as solid faint lines. As with Fig.~\ref{Mass_Functions}, the shaded areas show the bootstrap error estimate on the actual EAGLE clustering. The bootstrap method is done on a galaxy basis, which is still adequate in this case as we are not trying to quantify the impact of sample (or cosmic) variance: the models use the same set of DM halos as the EAGLE data, with only the stellar mass of their host galaxies possibly differing.
The correlations functions from each of our six models and for each stellar mass bin are shown as solid faint lines. Fig.~\ref{CorrelationBins} shows that our correlation functions agree within errors with the ones from EAGLE, which suggests that our models assign galaxy masses in a way that is sufficiently accurate to reproduce the stellar mass clustering of central galaxies up to $10 h^{-1} \text{Mpc}$. It is also noticeable that the scatter on the correlation functions from our methodology is smaller that the one from bootstrap errors. Hence to be able to differentiate between the models a significantly larger simulation volume would be needed.

Hydrodynamical N-body simulations that model both the dark matter and the baryonic component of the universe are computationally expensive. This limits the volumes in which they can be computed to a few $(100 \text{Mpc})^3$. Our models are informed by the physical processes relating the stellar mass of a galaxy and its host DM halo. Therefore, by populating DM-only simulations in larger volumes, our models could provide new tests of the hydrodynamical physics on larger scales than the ones permitted by direct comparisons with hydrodynamical simulations. The fact that we can reproduce accurately with our models both the stellar mass and the correlation functions of EAGLE, suggests that this approach is promising for populating DM only simulations. 

\section{Discussion and Conclusions}
\label{conclusions}

There is a well-known correlation between the stellar mass of a galaxy and the dark matter of its host halo (SMHM relation). However, this relation has significant scatter, which suggests that other properties are significant at determining the stellar mass of a halo. The halo mass evolution history and the specific angular momentum have both been proposed to be correlated with this residuals.

We use a sparse regression methodology to model the governing equations relating the stellar mass of central galaxies to the properties of their host dark matter halos. This method builds accurate and explainable models without needing much physical knowledge of the processes that determine the stellar mass of a galaxy from the halo properties of its host.
In sparse regression methods, the lack of physical knowledge is substituted by large numbers of free parameters, where each parameter models different behaviours of the dark matter halo properties. A LASSO algorithm is used to optimize solutions. This method heavily penalizes the number of surviving parameters so that as few as possible are selected without losing accuracy. Here we have modified the form of the LASSO algorithm to be more efficient when combined with a gradient descent minimizer. This is achieved by including a regularisation term that smooths out discontinuities in the gradient that are present in standard LASSO when parameters are close to zero. This smoothing is characterized by a parameter $\epsilon$ that limits how close to zero coefficients need to get before being discarded by the algorithm. We also modify the method by which the minimizer decides which path to follow in such a way that we find performance gains in large dimensional spaces.

The size of the penalty is determined by the parameter $\lambda$, which is optimized using a k-fold methodology with $k=10$. We use the one-standard-error rule to select a value of $\lambda$ that is larger than the best-fit and therefore builds a slightly less accurate model with fewer free parameters and therefore with more explainability. 

The data that we use to build our models with comes from the EAGLE simulation. However, we emphasize that this method should be able to be calibrated against any simulation with similar results. We use a sample of 9521 central galaxies from the 100\,cMpc box EAGLE suite of hydrodynamical simulations. The dark matter properties are read from a DM only simulation with the same initial conditions as our hydrodynamical simulation. The simulations are matched with each other in such a way that a pair is found for $99\%$ of the DM halos. 

We build four different models that differ by the independent parameters chosen to model the galaxy stellar mass. 
In the first instance, we consider two distinct model setups:
(i) the mass ratio model uses the ratio between the mass of a halo at a redshift $z$ and that at $z=0$ to parametrise the mass history of the host halo; (ii) the formation criterion model uses the redshift at which a halo formed a specific percentage of its mass. For both models we include all linear, quadratic and cubic correlations of our independent variables as free parameters of the fits. 
Then we consider two additional models by extending the two previous models to include parameters related to the specific angular momentum ($\vec{\mathbf{j}}$) history of the halos. More specifically, we consider parameters that characterize both the magnitude and the direction of the specific angular momentum vector, and vary the radius of the DM halo over which to measure the magnitude of $\vec{\mathbf{j}}$.  Due to computational restrictions, we include only linear terms of the free parameters related to $\vec{\mathbf{j}}$.   

The computational time of our minimization is correlated with the value of $\epsilon$: a very large value would result in a very fast computational time, but it would be hard also to distinguish useful parameters from those that should be discarded. In Fig.~\ref{epsilon_evo} we show that a value of $\epsilon=1 \times 10^{-3}$ selects the same coefficients as slower and more accurate runs without being too computationally expensive. Some input parameters are correlated with each other, for example, the mass ratio $(M_z/M_0)$ at a given redshift and that at a neighbouring redshift slice. In principle, our answers could be susceptible to the starting point of the minimizer; however, we show in Figs.~\ref{DiffIP} that neither the explainability nor the accuracy of the model changes significantly between runs with different starting points. We show in Fig.~\ref{DiffHold} that models trained on different subsets of the same data arrive at equivalent models.

Our algorithm did not select any  angular momentum parameters for either model that included specific angular momentum parameters. 

In fact, all the differences between these two models and their equivalent ones without angular momentum parameters are consistent with variations in our methodology. This suggests that any correlation between the linear terms of the angular momentum of a host halo and the residual of the SMHM relation is the consequence of correlations between the mass history of the halo and the history of its angular momentum.  
Given that model the formation criterion model is slightly simpler than the mass ratio model, we conclude that the formation criteria parameters, $\mathbf{FC}_p$, are slightly more efficient at summarizing the halo mass evolution information than the mass ratios $(M_z/M_0)$.

The formation criterion model is more accurate, although more complex, than models that include only halo mass terms, or models that also include a linear dependence on a single formation time. The improvement is, however, modest. Including a single linear formation time explains $32\%$ of the residual variance, while the full models improves this by a further $10\%$. If greater simplicity is required, this can be achieved (at the expense of accuracy) by increasing the penalty hyperparameter,  $\lambda$. The resulting model prefers to select terms that more closely characterise the formation history of the halo rather than terms the mix formation time and halo mass, however.

A subset of our surviving terms can be combined into a polynomial of $M_0$ and is therefore a model of the SMHM relation.  Other subsets of surviving terms can be combined into polynomials of either $\mathbf{FC}_p$ or  $M_z/M_0$ (depending on the parametrization of the halo mass evolution history) and therefore model the assembly bias. Terms of the shape $M_0 \times \mathbf{FC_p}$ (or $M_0 \times M_z/M_0$) add a significant correction to very small or very large halos. Our models suggest that a single formation time is not enough to model the variation in the SMHM relation, and that a better approach is to include the times at which different percentages of the mass have been formed. This is reflected in our model by the similar contribution of terms of the form $\mathbf{FC}_p$ for all $p$ in $p=[20,30,50,70,90]$. Our model also suggests that the relation between the stellar mass and the formation times is not the same for all galaxies, but it depends on the halo mass at $z=0$.

We have shown how the stellar mass function (SMF) of our model compares to that of EAGLE central galaxies. They agree well within the bootstrap errors at most stellar mass values, except around $\log_{10}(M^*/M_\odot)=9.0$ and $\log_{10}(M^*/M_\odot)=10.5$. The difference at lower stellar mass could be explained by selection effects given that our model includes a cut on halo mass that could affect the prediction of the lower stellar masses. At $\log_{10}(M^*/M_\odot)=10.5$ on the other hand, the differences between the values predicted by our model and EAGLE are not symmetric around the mean. This suggests that the remaining residual of our model might be correlated with variables that have not been explored by our model. This could be either higher-order correlations of our current variables, DM variables that we have not considered yet, or the stochastic effects of the baryon physics affecting the stellar mass of the galaxy. These will be studied in further extensions of the model.
We have also shown that the correlation function of EAGLE galaxies split by stellar mass is preserved in our models within the quoted bootstrap errors at all scales considered. 

The fact that we can reproduce both the stellar mass and the correlation function of EAGLE accurately suggests that this method could be used to populate DM only simulations in larger volumes in a way that preserves these statistics. Our models are informed by the physical process that relates the stellar mass of a galaxy with the evolutionary and present properties of its host DM halo. Therefore DM only simulations that are populated using our methodology can provide tests of this physics on volumes where hydrodynamical simulations are prohibitively expensive to run.  
So far, however, our method has only been applied to central galaxies. Satellite galaxies in general have a weaker SMHM relation than halos. This is a consequence of satellites being subjected to processes like tidal stripping and heating. These processes modify the mass of subhalos and the galaxies they contain, meaning that the stellar mass of a satellite halo is different from what one would expect when comparing it with halos that were not stripped. By adapting our methodology to account for the more complex evolution of the satellite halo mass, for example by adding maximum progenitor masses to our list of variables, it should be possible to model the stellar mass of satellite galaxies as well. However, running our methodology with satellite galaxies would require to use a larger set of free parameters and a larger data set, as there are many more satellite galaxies than central galaxies. Therefore we should explore methods to optimize our minimization without losing reliability. One approach could be to use methods like principal component analysis to transform free parameters into a parameter space where they are uncorrelated.  However, this might transform free parameters into inputs that are harder to interpret and might reduce the explainability of our results. These ideas will be explored in future iterations of this work.

\section*{Acknowledgements}
We thank Jorryt Matthee for helping us with the comparisons to his model, and for his invaluable insight in galaxy-halos relations.
MIL is supported by a PhD Studentship from the Durham Centre for Doctoral Training in Data Intensive Science, funded by the UK Science and Technology Facilities Council (STFC, ST/P006744/1) and Durham University. MIL, RGB, PN and SMC acknowledge support from the Science and Technology Facilities Council (ST/P000541/1 and ST/T000244/1).
This work used the DiRAC@Durham facility managed by the Institute for Computational Cosmology on behalf of the STFC DiRAC HPC Facility (www.dirac.ac.uk). The equipment was funded by BEIS capital funding via STFC capital grants ST/K00042X/1, ST/P002293/1 and ST/R002371/1, Durham University and STFC operations grant ST/R000832/1. DiRAC is part of the National e-Infrastructure.

\section{Data Availability}

 The data used in this work can be shared if requested from the authors. The data from the EAGLE simulations has been publicly released, see  \cite{2016A&C....15...72M}.



\bibliographystyle{mnras}
\bibliography{SMHM}


\bsp	
\label{lastpage}
\end{document}